\documentclass[useAMS,usenatbib]{mn2e}

\usepackage{graphicx}
\usepackage{txfonts}
\usepackage{amssymb,alltt}
\usepackage{color}
\usepackage{hyperref}
\usepackage{supertabular}

\hypersetup{colorlinks, linkcolor=blue}

\title[SN Refsdal in MACSJ1149.6+2223]{\textit{Hubble Frontier Fields}: Predictions for the Return of \emph{SN Refsdal} with the MUSE and GMOS Spectrographs}
\author[Jauzac et al. 2015]
{M. Jauzac,$^{1,2,3}$\thanks{E-mail:
mathilde.jauzac@durham.ac.uk} J. Richard,$^{4}$ M. Limousin,$^{5}$ K. Knowles,$^{3}$ G. Mahler,$^{4}$ G. P. Smith,$^{6}$
\newauthor
J.-P. Kneib,$^{7,5}$ E. Jullo,$^5$ P. Natarajan,$^{8}$ H. Ebeling,$^{9}$ H. Atek,$^{8}$ B. Cl\'ement,$^{4}$ D. Eckert,$^{10}$ 
\newauthor
E. Egami,$^{11}$ R. Massey,$^{1,2}$ M. Rexroth,$^{7}$
\\
\\
$^{1}$Centre for Extragalactic Astronomy, Department of Physics, Durham University, Durham DH1 3LE, U.K.\\
$^{2}$Institute for Computational Cosmology, Durham University, South Road, Durham DH1 3LE, U.K.\\
$^{3}$Astrophysics and Cosmology Research Unit, School of Mathematical Sciences, University of KwaZulu-Natal, Durban 4041, South Africa\\
$^{4}$CRAL, Observatoire de Lyon, Universit\'e Lyon 1, 9 Avenue Ch. Andr\'e, 69561 Saint Genis Laval Cedex, France\\
$^{5}$Laboratoire d'Astrophysique de Marseille - LAM, Universit\'e d'Aix-Marseille $\&$ CNRS, UMR7326, 38 rue F. Joliot-Curie, 13388 Marseille Cedex 13, France\\
$^{6}$School of Physics and Astronomy, University of Birmingham, Birmingham, B15 2TT, England\\
$^{7}$Laboratoire d'Astrophysique, Ecole Polytechnique F\'ed\'orale de Lausanne (EPFL), Observatoire de Sauverny, CH-1290 Versoix, Switzerland\\
$^{8}$Department of Astronomy, Yale University, 260 Whitney Avenue, New Haven, CT 06511, USA\\
$^{9}$Institute for Astronomy, University of Hawaii, 2680 Woodlawn Drive, Honolulu, Hawaii 96822, USA\\
$^{10}$Astronomy Department, University of Geneva, 16 ch. d'Ecogia, CH-1290 Versoix, Switzerland\\
$^{11}$Steward Observatory, University of Arizona, 933 North Cherry Avenue, Tucson, AZ, 85721, USA} 

\begin{document}

\date{Accepted XXXX. Received 2015 September XX; in original form 2015 December 20}

\pagerange{\pageref{firstpage}--\pageref{lastpage}} \pubyear{2015}

\maketitle

\label{firstpage}

\begin{abstract}
We present a high-precision mass model of the galaxy cluster MACSJ1149.6+2223, based on a strong-gravitational-lensing analysis of \emph{Hubble Space Telescope Frontier Fields} (HFF) imaging data and spectroscopic follow-up with Gemini/GMOS and VLT/MUSE.
Our model includes 12 new multiply imaged galaxies, bringing the total to 22, comprised of 65 individual lensed images. Unlike the first two HFF clusters, Abell 2744 and MACSJ0416.1$-$2403, MACSJ1149 does not reveal as many multiple images in the HFF data. 
Using the \textsc{Lenstool} software package and the new sets of multiple images, we model the cluster with several cluster-scale dark-matter halos and additional galaxy-scale halos for the cluster members.
Consistent with previous analyses, we find the system to be complex, composed of five cluster-scale halos.  Their spatial distribution and lower mass, however, makes MACSJ1149 a less powerful lens. Our best-fit model predicts image positions with an \emph{RMS} of 0.91$\arcsec$.
We measure the total projected mass inside a 200~kpc aperture as ($1.840\pm 0.006$)$\times 10^{14}$M$_{\odot}$, thus reaching again 1\% precision, following our previous HFF analyses of MACSJ0416.1$-$2403 and Abell 2744. 
In light of the discovery of the first resolved quadruply lensed supernova, \textit{SN Refsdal}, in one of the multiply imaged galaxies identified in MACSJ1149, we use our revised mass model to investigate the time delays and predict the rise of the next image between November 2015 and January 2016.

\end{abstract}



\section{Introduction}
\label{intro}
Since the discovery of the first giant arcs \citep[][in Abell 370]{soucail88}, gravitational lensing has been recognized as one of the most powerful tools to understand the evolution and assembly of structures in the Universe. Gravitational lensing allows us to measure the dark-matter content of the lenses, free from assumptions regarding their dynamical state \citep[for reviews, see e.g.\ ][]{bible,2010RPPh...73h6901M,KN11,hoekstra13}, as well as to spatially resolve the lensed objects themselves \citep[][]{richard10,smith09,rau14}. Massive galaxy clusters are ideal ``cosmic telescopes'' and generate high magnification factors over a large field of view \citep[][]{ellis01,kneib04a}. Their importance for the study of both clusters and the distant Universe lensed by them is apparent from ambitious programs implemented with the \textit{Hubble Space Telescope} (HST), like the \textit{Cluster Lenses And Supernovae with Hubble} \citep[CLASH, PI :Postman;][]{postman12} multi-cycle Treasury project, the \textit{Grism Lens-Amplified Survey from Space} program \citep[GLASS, PI: Treu;][]{schmidt14}, and the recent \textit{Hubble Frontier Fields} (HFF) Director's initiative\footnote{http://www.stsci.edu/hst/campaigns/frontier-fields/}.  

The galaxy cluster studied in this paper, MACSJ1149.6+2223 (MACSJ1149 hereafter), at redshift $z=0.544$ (R.A: +11:49:34.3, Decl.: +22:23:42.5), was discovered by the MAssive Cluster Survey \citep[MACS,][]{ebeling01,ebeling07}.
The first strong-lensing analyses of MACSJ1149 were published by \cite{smith09}, \cite{zitrin09a}, and \cite{zitrin11}, based on shallow HST data (GO-9722, PI: Ebeling) taken with the \textit{Advanced Camera for Surveys} (ACS), and revealed one of the most complex cluster cores known at the time.
MACSJ1149 stands out among other massive, complex clusters not only by virtue of its relatively high redshift, but also for it hosting a spectacular lensed object, a triply-lensed face-on spiral at $z=1.491$ \cite[][]{smith09}.
The system was selected as a target for the CLASH program and thus observed with both ACS and the \textit{Wide Field Camera 3} (WFC3) across 16 passbands, from the UV to the near-infrared, for a total integration time of 20 HST orbits, leading to the discovery of a lensed galaxy at $z=9.6$, observed near the cluster core \citep[][]{zheng12}, and the publication of a revised strong-lensing analysis by \cite{rau14}.

More recently, MACSJ1149 was selected as one of the six targets for the HFF observing campaign. Combining the lensing power of galaxy clusters with the high-resolution of HST and allocating a total of 140 HST orbits for the study of each cluster, the HFF initiative aims to  probe the distant and early Universe to an unprecedented depth of $mag_{\mathrm{AB}}\!\sim\!29$ in seven passbands (3 with ACS, 4 with WFC3).
In a coordinated multi-team effort, mass models\footnote{http://archive.stsci.edu/prepds/frontier/lensmodels/} of all six HFF cluster lenses were derived from pre-HFF data (CLASH data in the case of MACSJ1149) to provide the community with a first set of magnification maps \citep[see in particular][]{johnson14,coe15,richard14}.
Deep HFF imaging of MACSJ1149 was obtained during Cycle 22.

In 2014, MACSJ1149 was observed with WFC3 between November 3$^{rd}$ and 20$^{th}$ as part of the GLASS  programme. In the resulting data, \cite{kelly15} discovered a new supernova (SN) within the multiply-imaged spiral galaxy discussed above, lensed into an Einstein Cross by a foreground cluster galaxy. Multiply-lensed SN have been predicted for years, but with a relatively low probability of detection \citep[e.g.][]{refsdal64,KP88}. Up to then, only candidates of such events had been reported \citep[][]{goobar09,quimby14,patel14}, making this new SN, named \textit{SN Refsdal} by their discoverers, the first secure case of a resolved multiply-lensed SN. We refer the reader to \cite{kelly15} for more details.
The discovery of \textit{SN Refsdal} led to revisions of the pre-HFF strong-lensing analysis of MACSJ1149 and allowed measurements of time delays as well as predictions for the time of appearance of the same SN event in another image of the multiply imaged spiral \citep{sharon15,oguri15,diego15}. 
More recently, other HFF analyses were presented in \cite{treu15}, \cite{grillo15} and \cite{kawamata15}, in good agreement with the analysis presented here.

In this paper, we present a revised and improved version of the mass model of MACSJ1149 by \cite{richard14}, taking advantage of the recent deep \textit{HFF} images of the system, as well as spectroscopic surveys of the cluster core with Gemini/GMOS and VLT/MUSE. 
We study the case of \textit{SN Refsdal} and compare our results with those obtained by \cite{sharon15}, \cite{oguri15}, \cite{diego15}, \cite{treu15}, \cite{grillo15}, and \cite{kawamata15}.

When quoting cosmology-dependent quantities, we adopt the $\Lambda$CDM concordance cosmology with $\Omega_{m} = 0.3$, $\Omega_{\Lambda} = 0.7$, and a Hubble constant $H_0 = 70$~km$\,$s$^{-1}\,$Mpc$^{-1}$. Magnitudes are quoted in the AB system.

\begin{figure*}
\begin{center}
\includegraphics[width=0.97\textwidth]{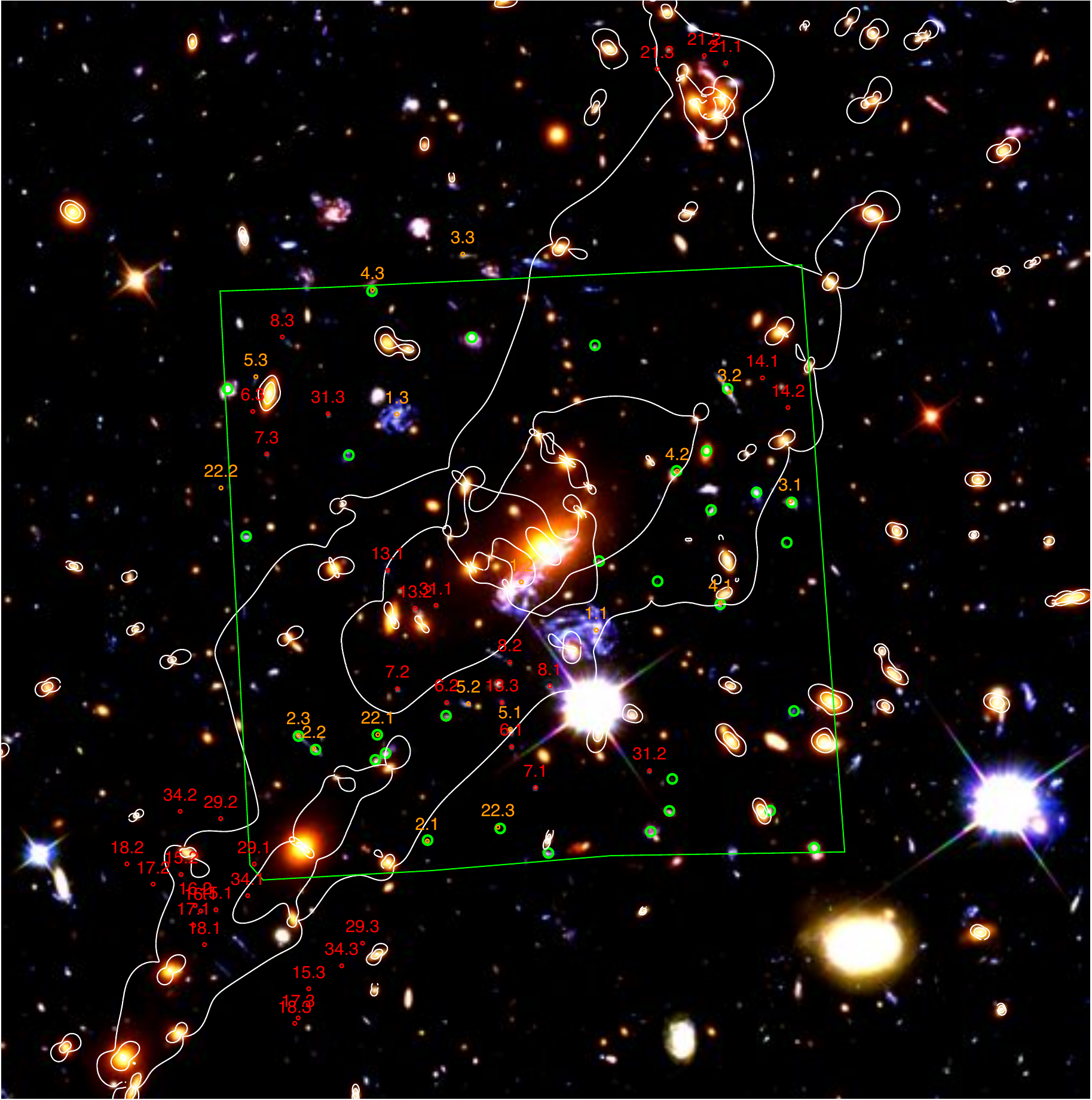}\\
\caption{Overview of all multiple-image systems used in this study. The most secure identifications, used to optimise the lens model in the \textit{image plane} (65 images) are shown in red; in orange we highlight the systems with a spectroscopic redshift from either GMOS or MUSE, with larger green circles highlighting the background sources with a MUSE redshift.
System \#1 is split into 24 individual sources at the same redshift, not labelled on the figure for clarity (see Table.~\ref{multiples} for their coordinates).
The underlying colour image is a composite created from HST/ACS images in the F814W, F606W, and F435W passbands. Critical lines at $z=1.49$, and $z=7.0$ are shown in white. The green rectangle highlights the VLT/MUSE field of view. The top left inset shows a close-up view of the Northern component of the cluster (Clump \#4 in Table~\ref{SLmodel_res}). North is up and East is left.}
\label{multiples}
\end{center}
\end{figure*}

\section{ Observations}
\label{observations}

\subsection{\textit{Hubble Frontier Fields} Data}
MACSJ1149 was observed for the \textit{HFF} campaign (ID: 13504, P.I: J. Lotz) with WFC3 between November 2014 and January 2015 in four filters, and with ACS between April and May 2015 in three filters. The discovery of \textit{SN Refsdal} in the GLASS data led to additional observations with WFC3, performed (and to be performed) at preset intervals between January and November 2015 (ID: 13790, PI: Rodney).
We used the self-calibrated data (version v1.0) with a pixel size of 0.03\arcsec provided by STScI\footnote{http://archive.stsci.edu/pub/hlsp/frontier/m1149/images/hst/}. These data combine all \textit{HST} observations of the cluster for total integration times corresponding to 25, 20.5, 20, and 34.5 orbits with WFC3 in the F105W, F125W, F140W, and F160W passbands, respectively, and to 18, 10, and 42 orbits with ACS in the F435W, F606W, and F814W filters, respectively, leading to a limiting magnitude of $mag_{\rm AB} = 29$, and thus a depth typical of ultra-deep field observations, for all seven filters. A composite HST/ACS colour image is shown in Fig.~\ref{multiples}.

\subsection{Spectroscopy with GMOS}
MACSJ1149 was observed with the GMOS spectrograph on Gemini North (ID program: GN-2010A-Q-8) during four nights between 
March 19$^{th}$ and April 20$^{th}$ of 2010. The seeing varied between 0.7\arcsec and 0.9\arcsec.  A single multi-object mask was used with a total of 20 slits 
covering multiple images, cluster members and other background galaxies identified from the 
HST images; the slit width was 1\arcsec. Observations with the B600 and R831 gratings provided a spectral resolution between 1500 at 650 nm and 3000 at 840 nm. 
A total of 40$\times$1050 second exposures were taken, equally split across four wavelength 
settings centred at 540, 550, 800, and 810 nm.

The GMOS spectroscopic data were reduced using the Gemini {\sc IRAF} reduction package (v. 1.1) 
to create individual calibrated 2D spectra of each slit and exposure. These were then aligned 
and combined using standard {\sc IRAF} recipes, and 1D spectra were extracted at the location 
of the sources of interest.

\subsection{Spectroscopy with MUSE}
The integral field spectrograph MUSE \citep{bacon10} on the \textit{Very Large Telescope} (VLT) observed the very central region of MACSJ1149 (green square in Fig.~\ref{multiples}) on February 14$^{th}$ and March 21$^{st}$ 2015 as part of the DDT program 294.A-5032(A) (PI: Grillo). The spectrograph's 1$\times$1 arcmin$^2$ field of view was rotated slightly to a position angle of 4 degrees in order to include the majority of the central multiple-image systems.
The seeing varied between 0.9 and 1.2\arcsec.

\begin{center}
\begin{figure}
\hspace*{-2mm}
\includegraphics[width=0.49\textwidth,angle=0.0]{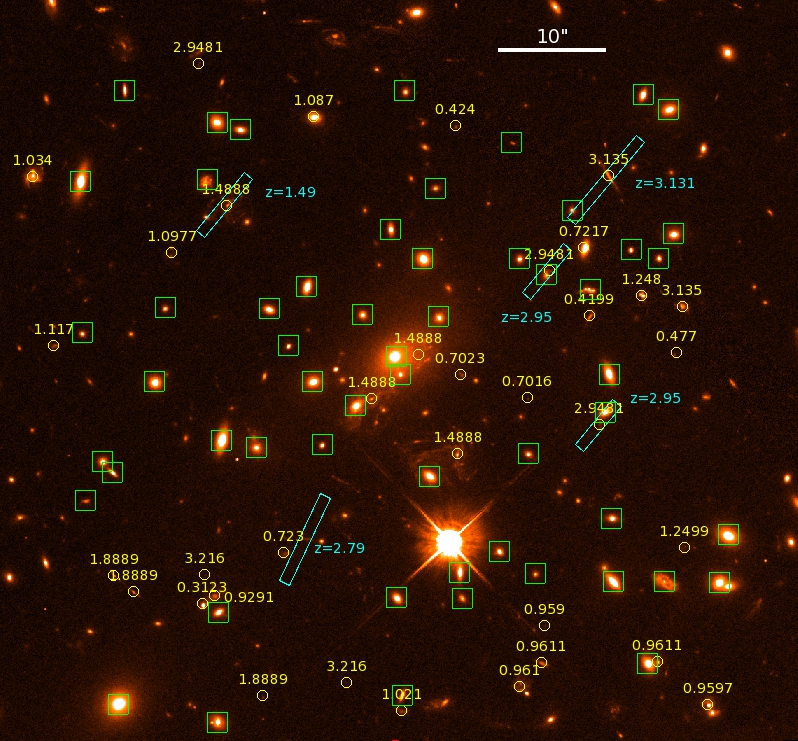}
\caption{HST/ACS F814W image of MACSJ1149 centred on the MUSE field of view. Cyan slits mark GMOS spectroscopic measurements; yellow circles show the sources with MUSE spectroscopic redshifts; and green squares highlight cluster members with a MUSE spectroscopic redshift as listed in Table~\ref{CM_muse}. We note the presence of a small group of galaxies at $z=0.96$ in the south-west corner.
}
\label{muse_FOV}
\end{figure}
\end{center}

For the analysis presented here, we combine ten exposures of 1440 seconds each that are publicly available from the ESO archive \citep[the complete observations were published while this manuscript was under review,][]{grillo15}. We reduced these data using version 1.1 of the MUSE data reduction pipeline (Weilbacher et al. \textit{in prep.}); selected results from the full data set (including proprietary exposures) are presented by \cite{karman15}.
We performed the basic calibrations (bias and flat-field corrections, wavelength and geometrical calibration) and applied a twilight and illumination correction to the data taken on each night. Flux calibration was performed using a standard star taken at the beginning of the night, and a global sky subtraction was applied to the pixel tables before a final resampling. The ten exposures were then aligned after adjustments for offsets measured from the centroid of the brightest star in the South of the field (Fig.~\ref{muse_FOV}). 

The final MUSE data cube has a spatial pixel scale of 0.2$\arcsec$ and covers the wavelength range 4750--9350~\AA\ at 1.25 \AA/pixel and a resolution of 1500--3000. Following  \citet{richard15}, we analysed this large dataset using two complementary approaches: we first extracted the 1D spectrum at the location of each of the sources detected in the HFF images and falling within the MUSE field of view, and then estimated all possible redshifts based on emission- and absorption-line features. 
In addition, we used narrow-band images created with customized software based on \textsc{Sextractor} \citep{BA96} to perform a blind search of the data cube for isolated emission lines not associated with continuum sources. We then merged the results from both approaches to generate a final MUSE redshift catalog. 

In total, we measured the redshift of 88 sources, including 57 cluster members between $z=0.513$ and $z=0.570$, and 27 background sources (some of them being multiply imaged, see Sect.~\ref{zspec_im}). Tables~\ref{CM_muse}, \ref{bkg_muse}, and \ref{fgd_muse} list coordinates and redshifts for cluster members, singly imaged background sources, and  foreground galaxies, respectively. The redshifts for multiple images are provided in Table~\ref{multipletable}.

\begin{center}
\begin{figure}
\hspace*{-2mm}
\includegraphics[width=0.49\textwidth,angle=0.0]{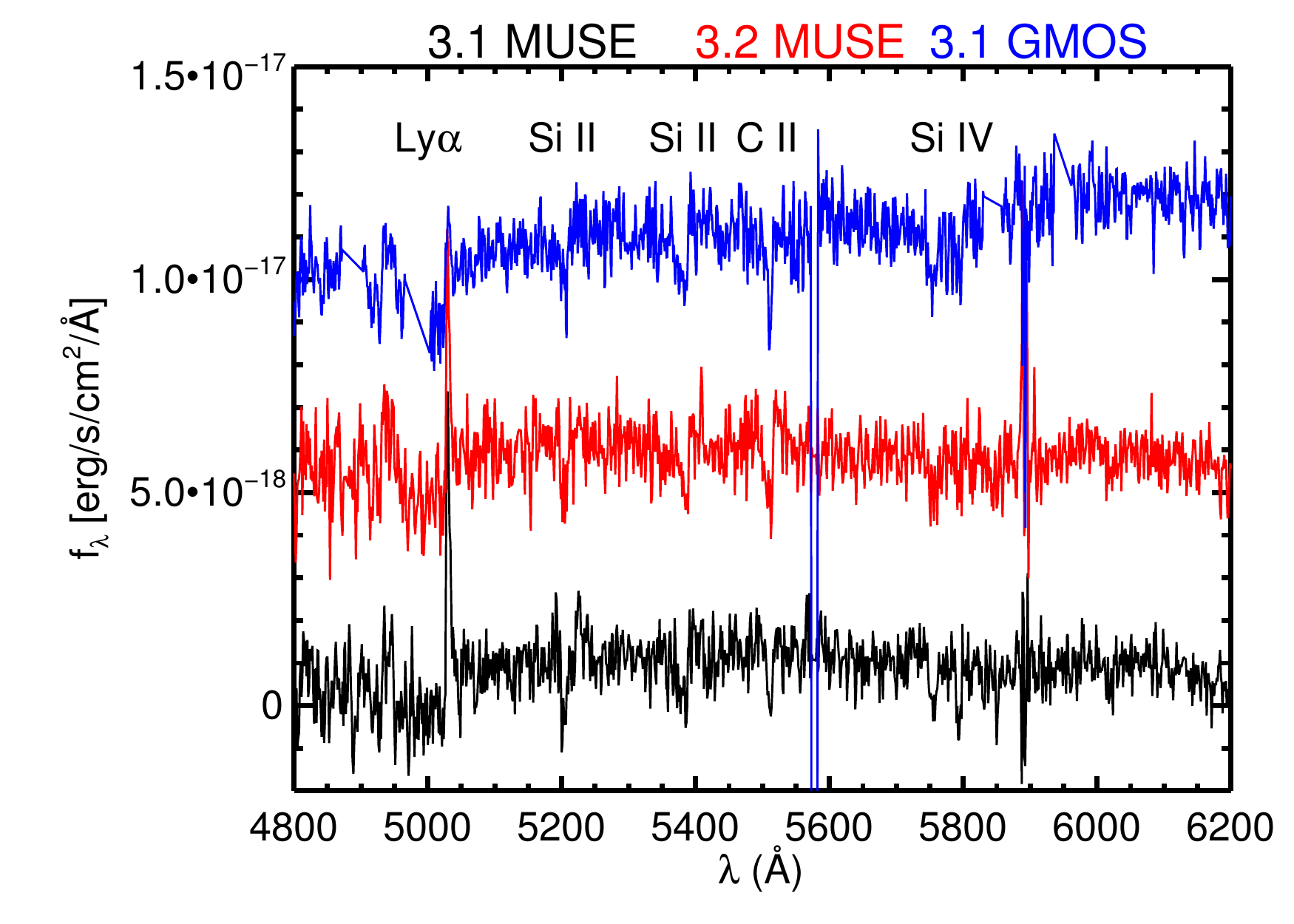}
\caption{Example of extracted GMOS and MUSE spectra for system 3, confirming a redshift $z=3.128$ from Lyman-$\alpha$ in emission, spectral break, and UV absorption lines. Individual spectra are offset vertically for clarity.}
\label{spectra}
\end{figure}
\end{center}

\begin{table}
\begin{center}
\caption{Catalogue of cluster members detected with VLT/MUSE observations.}
\label{CM_muse}
\begin{tabular}[h!]{cccc}
\hline
\hline
\noalign{\smallskip}
\textbf{ID} & \textbf{R.A.} & \textbf{Decl.} & $z_{\rm spec}$\\
\hline
\hline
1 & 177.39548 & 22.404037 & 0.5133\\
2 & 177.39328 & 22.400253 & 0.5134\\
3 & 177.39859 & 22.398064 & 0.5264\\
4 & 177.39096 & 22.401691 & 0.5272\\
5 & 177.40628 & 22.405381 & 0.5277\\
6 & 177.40358 & 22.396369 & 0.5307\\
7 & 177.40745 & 22.399136 & 0.5307\\
8 & 177.39121 & 22.392715 & 0.531\\
9 & 177.40261 & 22.396186 & 0.5315\\
10 & 177.39287 & 22.397096 & 0.5322\\
11 & 177.40546 & 22.397881 & 0.5327\\
12 & 177.40121 & 22.400339 & 0.5327\\
13 & 177.39181 & 22.405281 & 0.5335\\
14 & 177.3911 & 22.404904 & 0.5335\\
15 & 177.40306 & 22.404389 & 0.5335\\
16 & 177.39846 & 22.405383 & 0.536\\
17 & 177.39854 & 22.389783 & 0.536\\ 
18 & 177.39965 & 22.399616 & 0.536\\ 
19 & 177.39139 & 22.401063 & 0.5365\\
20 & 177.39502 & 22.39602 & 0.5365\\
21 & 177.3938 & 22.402294 & 0.5385\\
22 & 177.40752 & 22.403047 & 0.5392\\
23 & 177.39686 & 22.392292 & 0.5398\\
24 & 177.40077 & 22.396256 & 0.5403\\
25 & 177.40104 & 22.397885 & 0.5403\\
26 & 177.40515 & 22.399789 & 0.5408\\
27 & 177.39581 & 22.393496 & 0.5408\\
28 & 177.39215 & 22.401282 & 0.5408\\
29 & 177.39869 & 22.392303 & 0.5411\\
30 & 177.3987 & 22.398519 & 0.5411\\
31 & 177.39452 & 22.400647 & 0.5416\\
32 & 177.39886 & 22.401818 & 0.5418\\
33 & 177.40014 & 22.394428 & 0.5425\\
34 & 177.39527 & 22.401054 & 0.5426\\
35 & 177.40171 & 22.398803 & 0.5428\\
36 & 177.40367 & 22.391933 & 0.5433\\
37 & 177.39483 & 22.392927 & 0.5436\\
38 & 177.39169 & 22.390611 & 0.5436\\
39 & 177.40225 & 22.399759 & 0.5436\\
40 & 177.38969 & 22.392704 & 0.5441\\
41 & 177.4069 & 22.39583 & 0.5441\\
42 & 177.39778 & 22.395445 & 0.5443\\
43 & 177.39269 & 22.394364 & 0.5453\\
44 & 177.39752 & 22.39955 & 0.5458\\
45 & 177.40663 & 22.395536 & 0.5466\\
46 & 177.4037 & 22.404578 & 0.5468\\
47 & 177.40737 & 22.394819 & 0.547\\
48 & 177.39797 & 22.401045 & 0.5471\\
49 & 177.40369 & 22.389101 & 0.5491\\
50 & 177.40645 & 22.389565 & 0.5496\\
51 & 177.39265 & 22.392733 & 0.5504\\
52 & 177.39693 & 22.39297 & 0.5511\\
53 & 177.39761 & 22.402875 & 0.5511\\
54 & 177.39275 & 22.398072 & 0.5519\\
55 & 177.38943 & 22.393942 & 0.5554\\
56 & 177.39983 & 22.397255 & 0.5609\\
57 & 177.40397 & 22.403094 & 0.567\\
\hline
\hline
%
\end{tabular}
\end{center}
\end{table}

\section{Multiply Imaged Systems}
\label{Mul_SL}

\subsection{HST identifications}
MACSJ1149 has been the subject of a number of strong-lensing analyses \citep[][]{smith09,zitrin09a,zitrin11,rau14,richard14,johnson14,coe15,sharon15,oguri15,diego15}, all of which were based on pre-HFF data except for \cite{diego15}, their work uses one third of the HFF data.
We started our search for multiple images guided by the mass model of \cite{richard14}.
This model incorporates 35 multiple images of 10 different lensed galaxies, three of which have spectroscopic redshifts from \cite{smith09}: systems \#1, \#2, and \#3, at $z=1.491$, $z=1.894$, and $z=2.497$, respectively. 

The new, deep HFF ACS and WFC3 images allow us to extend this set of multiple images. To this end, we followed \cite{jauzac14} and \cite{jauzac15b} and first computed the cluster's gravitational-lensing deflection field that describes the mapping of images from the image plane to the source plane, on a grid with a spacing of 0.2 arcsec/pixel. 
Since the transformation scales with redshift as described by the distance ratio $D_{LS}/D_{OS}$, where $D_{LS}$ and $D_{OS}$ are the distances between the lens and the source, and the observer and the source, respectively, it is only computed once, thereby enabling an efficient lens inversion. We then compute the critical region at redshift $z=7$ and limit our search for multiple images in the ACS data to this area (white contours in Fig.~\ref{multiples}).  
Careful searches, combined with visual scrutiny and confirmation, revealed 12 new multiply imaged systems, bringing the total number of multiple images identified in MACSJ1149 to 65, involving 22 different multiply imaged galaxies (Fig.~\ref{multiples} and Table~\ref{multipletable}), which leads to considerable tighter constraints on the mass model of the cluster.
Although a significant improvement over the pre-HFF statistics, this number of new systems is disappointing compared to how many were discovered in the first two HFF clusters, MACSJ0416.1--2403 and Abell 2744; we discuss this issue in Sect.~\ref{lenspower}. 

As one of the main goals of our analysis is to measure precise time delays prompted by the discovery of  \textit{SN Refsdal}, we followed \cite{rau14} and decomposed System \#1, the SN host galaxy, into 24 features, selected as the brightest components of the spiral (see the bottom part of Table~\ref{multipletable} for their coordinates). We also added the four images of \textit{SN Refsdal} located in image 1.2 and labelled S1, S2, S3, and S4, following the same notation as \cite{kelly15}, \cite{sharon15} and \cite{oguri15}.

In order to test the reliability of our multiple-image identifications, we computed a flux-$\chi^2$ statistic to quantify the similarity of the photometry in each pair of images within a given system:

 $\chi^{2}_{\nu}=\frac{1}{N-1} \min_{\alpha}\left(\sum\limits_{i=1}^{N} \frac{ ( f_{i}^{A}- \alpha f_{i}^{B})^{2} } {{\sigma_{i}^{A}}^{2}+ \alpha^{2} {\sigma_{i}^{B}}^{2} }\right)$

where $f_i$ and $\sigma_i$ are the fluxes and errors in filter $i$, $N$ is the total number of filters, and $\alpha$ is the minimisation factor rescaling both SEDs.
As shown by Mahler et al.\ (in prep.) this statistic quantifies the probability of two images originating from the same source.  

Flux measurements were derived from isophotal magnitudes measured with \textsc{SExtractor} \citep{BA96}, and the fluxes of oversegmented multiple images were combined into a total flux per multiple image. The corresponding magnitudes are presented in Table \ref{multipletable}. Combining all HFF filters, we find acceptable values for $\chi^2$ ($\sim1-3$) for almost all images, with slightly high values typically being observed for sources whose photometry is compromised by bright nearby sources. 

\begin{center}
\begin{figure}
\hspace*{-2mm}
\includegraphics[width=0.159\textwidth,angle=0.0]{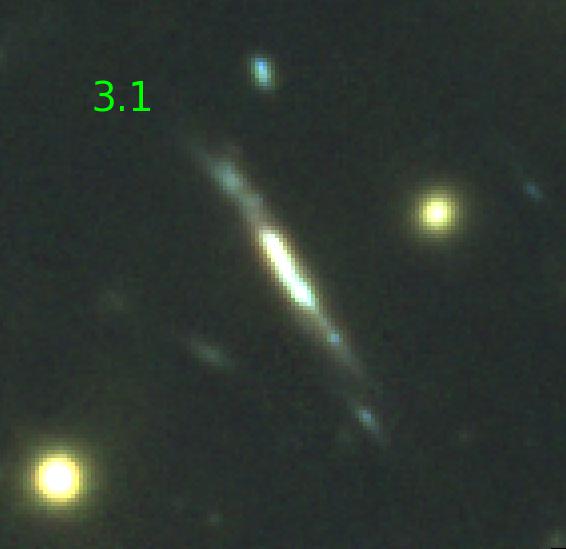}
\includegraphics[width=0.159\textwidth,angle=0.0]{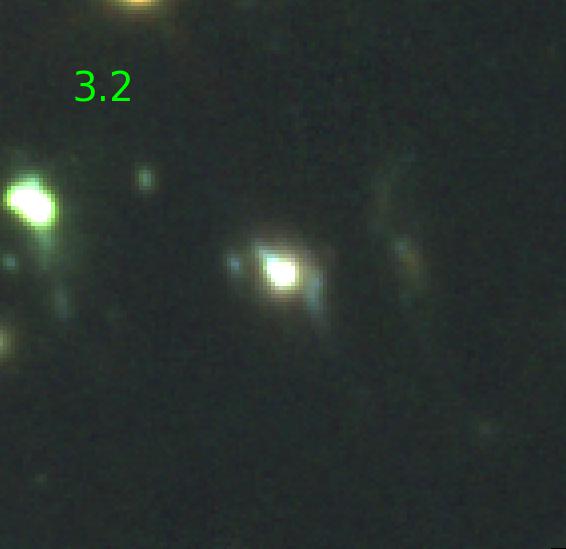}
\includegraphics[width=0.159\textwidth,angle=0.0]{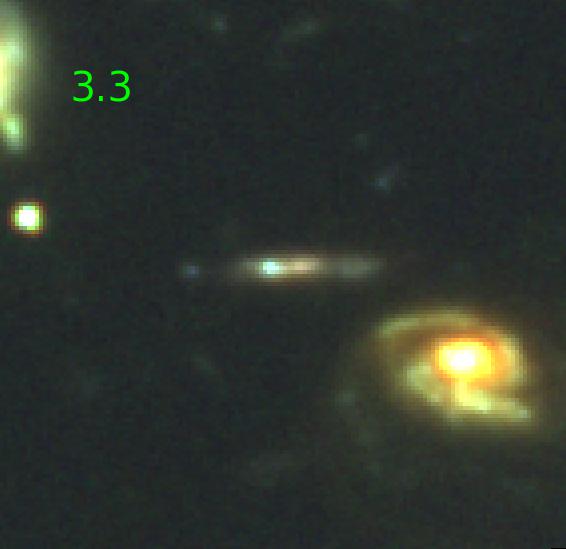}\\
\hspace*{-2mm}
\includegraphics[width=0.159\textwidth,angle=0.0]{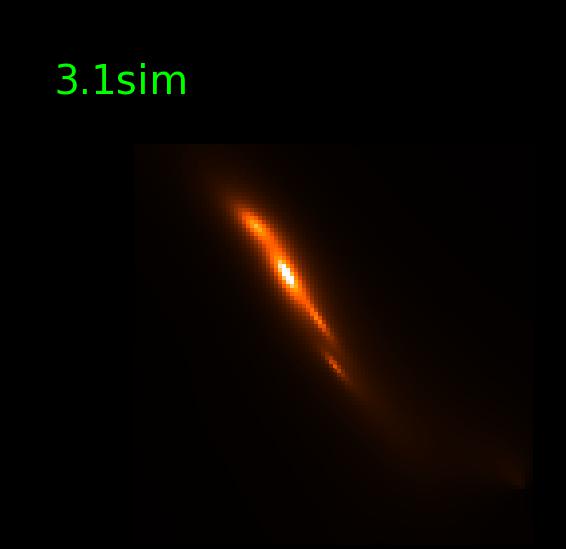}
\includegraphics[width=0.159\textwidth,angle=0.0]{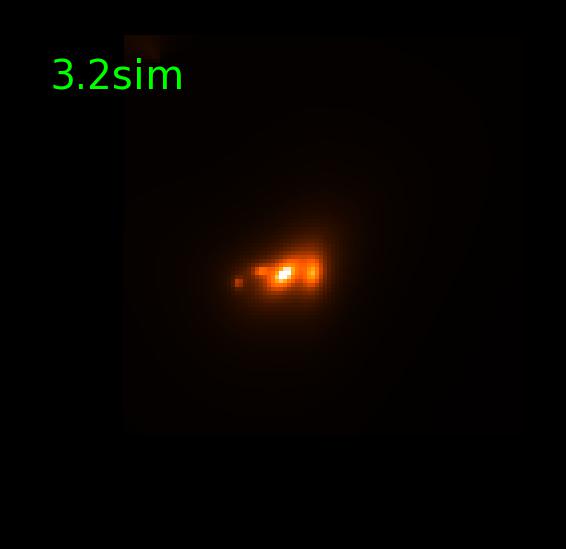}
\includegraphics[width=0.159\textwidth,angle=0.0]{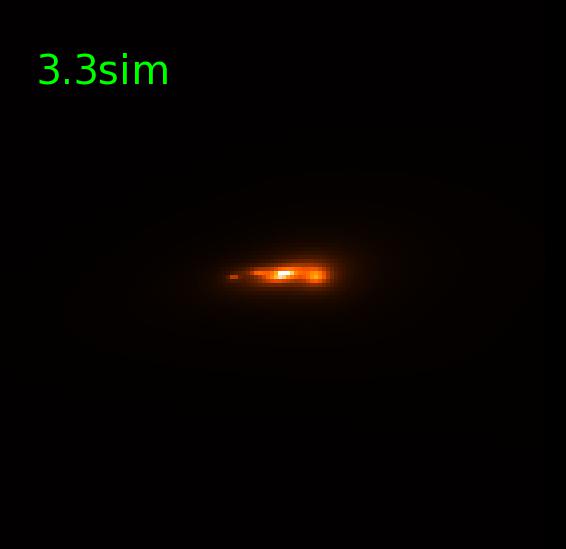}\\
\caption{\textit{Top panels:} HST ACS/WFC3 colour images of system 3 (F814W, F105W and F160W filters as RGB). \textit{Bottom panels:} Predicted images from a source matching the morphology of image 3.1.}
\label{sys3}
\end{figure}
\end{center}

\begin{center}
\begin{figure}
\hspace*{-2mm}
\includegraphics[width=0.49\textwidth,angle=0.0]{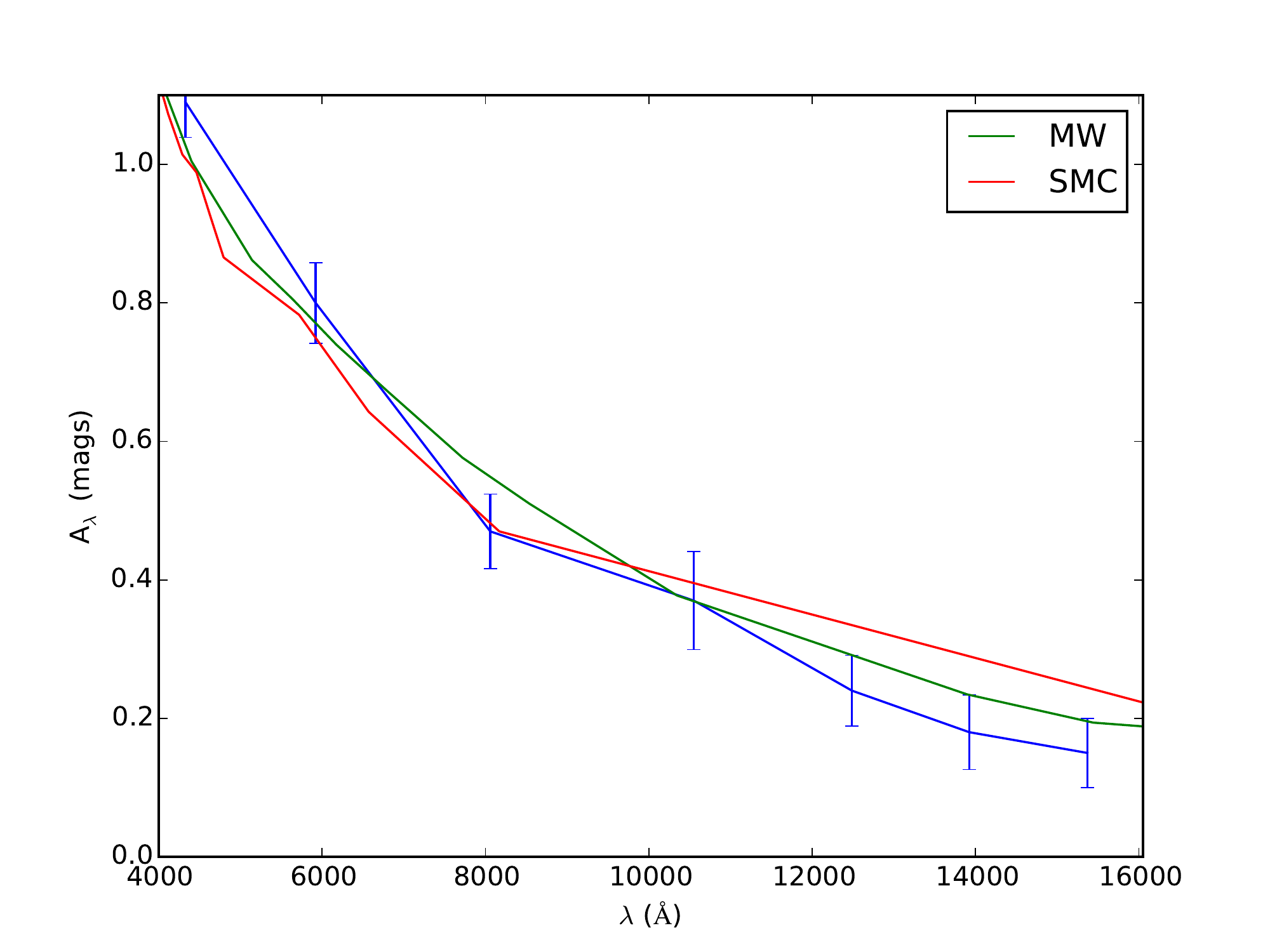}
\caption{Extinction estimated from a comparison of image 3.3 and image 3.1, and models assuming a Milky-Way or SMC extinction law, with attenuations $A_V=0.51$ and $A_V=0.47$ respectively.}
\label{red}
\end{figure}
\end{center}

A notable exception is image 3.3, already found in the pre-HFF images, which features a very high $\chi^2$ value (56), although it seems to be the most plausible counter-image of system 3 based on predictions of both its position and morphology derived from a lensing model constrained with 3.1 and 3.2 only. Fig.~\ref{sys3} shows the three images of system \#3 in composite HST ACS/WFC3 colour images (top panel), as well as the predicted images (monochrome), simulated based on the morphology of image 3.1. 
The predicted location and morphology of images 3.2 and 3.3 closely match those of the real data; 
however, the colour of image 3.3 is significantly reddened compared to images 3.1 and 3.2, 
thus producing the aforementioned large $\chi^2$ value. Removing the differential amplification between images 3.1 and 3.3, we find the magnitude differences in all filters to follow a typical reddening curve (Figure \ref{red}), which can be easily modelled by a Milky-Way (MW, \citealt{Allen76}) or SMC \citep{Bouchet,Prevot} extinction curve, with typical values of $A_V=0.51$ (MW) or $A_V=0.47$ (SMC), 
if we assume that the extinction occurs in the cluster at $z=0.54$. Dust extinction has been previously reported in the outskirts of clusters (e.g., \citealt{chelouche07}). Alternatively, we cannot rule out dust extinction by an intervening galaxy in the foreground or background of the cluster. 
For instance, the background spiral to the lower right of image 3.3 (Fig.~\ref{sys3}) is a good candidate. We derive a photometric redshift $z=1.4$ and a best-fit extinction $A_V=1.0$ from the public CLASH photometric catalogs \citep[][]{postman12}\footnote{https://archive.stsci.edu/missions/hlsp/clash/macs1149/catalogs/hst/} using the {\sc Hyperz} photometric redshift software \citep{bolzonella00}. Image 3.3 is located 12~kpc from the center of this galaxy in the source plane at $z=1.4$; at such a high impact parameter the lower extinction in image 3.3 can thus be expected.

\subsection{Redshift Constraints}
\label{zspec_im}
\begin{center}
\begin{figure}
\hspace*{-2mm}
\includegraphics[width=0.49\textwidth,angle=0.0]{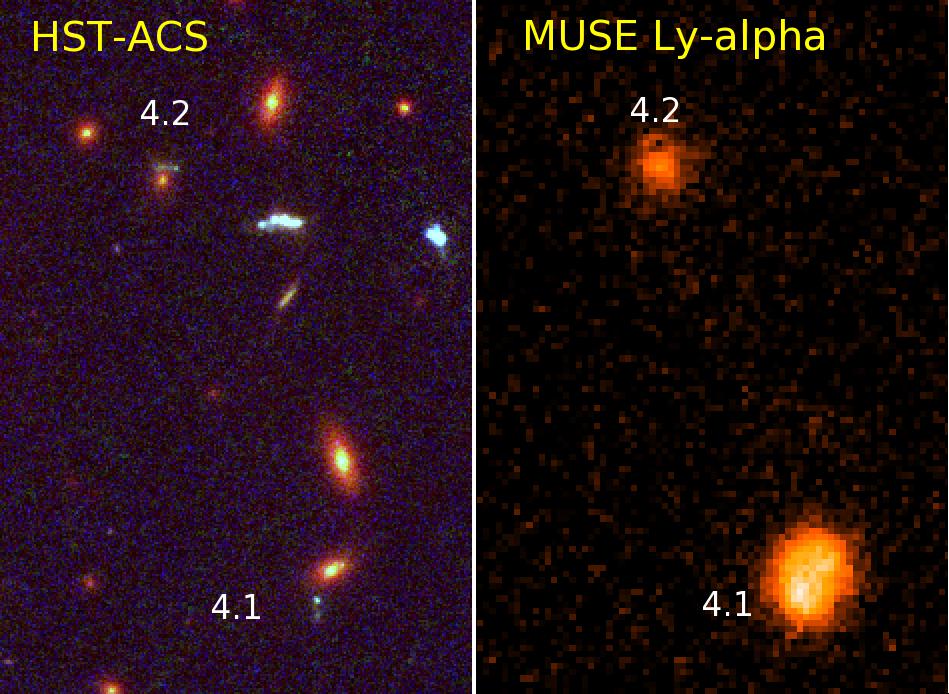}
\caption{Close-up view of System \#4. The left panel highlights images 4.1 and 4.2 in the HST/ACS image. The right panel shows the Lyman-$\alpha$ emission in the same area as observed with MUSE.}
\label{sys4}
\end{figure}
\end{center}

The spectroscopic observations described in Sect.~\ref{observations} allowed us to confirm newly identified multiple-image systems, as well as to correct earlier spectroscopic redshifts measurements.
Compared to \cite{smith09}, we revise the spectroscopic redshift of system \#3 to $z=3.128$, based on the deeper GMOS data that clearly show Lyman-$\alpha$ in emission as well as an associated spectral Ly-$\alpha$ break in the continuum (Fig.~\ref{spectra}). The revised redshift of system \#3 is also confirmed with MUSE for images 3.1 and 3.2. 
We also measure secure GMOS spectroscopic redshifts for system \#4, $z=2.95$, for system \#5, $z=2.79$, and system \#9, $z=0.981$, from strong Lyman-$\alpha$ and $[OII]$ emission lines, respectively. MUSE observations add to these findings by revealing extended emission around image 4.1, producing a Lyman-$\alpha$ Einstein ring around the very close cluster member (Fig.~\ref{sys4}). MUSE observations also confirm the redshift of the HFF multiple-image system \#22 as $z=3.216$, again from strong Lyman-$\alpha$ emission. Finally, we slightly revise the redshift of the system \#1, the face-on spiral, to $z=1.4888$, based on the total IFU spectrum of this object \citep[see also ][]{yuan11}.

In addition to measuring the redshifts of known multiple-image systems, we used the spectroscopic redshifts measured from GMOS and MUSE data also to investigate the possible multiplicity of other background sources (see second part of Table~\ref{bkg_muse}). 
All of these are predicted by our best mass model (Sect.~\ref{SLmass}) to be single images, including a small group of 11 galaxies at $0.95<z<1.3$ within the MUSE field of view (Fig.~\ref{muse_FOV} and Table~\ref{bkg_muse}). This test allowed us to reject potential new multiple images and to confirm the validity of our strong-lensing analysis.

Table~\ref{multipletable} lists the coordinates, redshifts (spectroscopic or predicted by our model),  F814W-band magnitudes, and magnifications predicted by our best-fit model, for all multiple images used in this work.  Magnitudes were measured using \textsc{Sextractor} \citep[][]{BA96}. 

\begin{table}
\begin{center}
\caption{Catalogue of singly-imaged background galaxies detected with VLT/MUSE observations.}
\label{bkg_muse}
\begin{tabular}[h!]{cccc}
\hline
\hline
\noalign{\smallskip}
\textbf{ID} & \textbf{R.A.} & \textbf{Decl.} & $z_{\rm spec}$\\
\hline
\noalign{\smallskip}
58 & 177.39503 & 22.397460 & 0.7016 \\
59 & 177.39691 & 22.398059 & 0.7023\\
60 & 177.39346 & 22.401332 & 0.7217\\
61 & 177.40183 & 22.393461 & 0.723\,\,\,\\
62 & 177.40377 & 22.392353 & 0.9291\\
63 & 177.39456 & 22.391586 & 0.959\,\,\,\\
64 & 177.39000 & 22.389538 & 0.9597\\
65 & 177.39525 & 22.390021 & 0.961\,\,\,\\
66 & 177.39141 & 22.390644 & 0.9611\\
67 & 177.39465 & 22.390637 & 0.9611\\
68 & 177.39854 & 22.389384 & 1.021\,\,\,\\
69 & 177.40885 & 22.403175 & 1.034\,\,\,\\
70 & 177.40100 & 22.404706 & 1.087\,\,\,\\
71 & 177.40495 & 22.401208 & 1.0977\\
72 & 177.40825 & 22.398792 & 1.117\,\,\,\\
73 & 177.39185 & 22.400103 & 1.248\,\,\,\\
74 & 177.39065 & 22.393606 & 1.2499\\
\hline
\hline
\end{tabular}
\end{center}
\end{table}

\begin{table}
\begin{center}
\caption{Catalogue of foreground galaxies detected with VLT/MUSE observations.}
\label{fgd_muse}
\begin{tabular}[h!]{cccc}
\hline
\hline
\noalign{\smallskip}
\textbf{ID} & \textbf{R.A.} & \textbf{Decl.} & $z_{\rm spec}$\\
\hline
\noalign{\smallskip}
  75 & 177.40410 & 22.392153 & 0.3123\\
  76 & 177.39331 & 22.399581 & 0.4199\\
  77 & 177.39704 & 22.404475 & 0.424\,\,\,\\
  78 & 177.39088 & 22.398611 & 0.477\,\,\,\\
  \hline
\hline
\end{tabular}
\end{center}
\end{table}

\section{Strong-Lensing Mass Model}
\label{SLmass}

\begin{table*}
\begin{center}
\begin{tabular}[h!]{cccccccc}
\hline
\hline
\noalign{\smallskip}
Clump  & $\Delta$ x  & $\Delta$ y & $e$ & $\theta$ & r$_{\mathrm{core}}$ (\footnotesize{kpc}) & r$_{\mathrm{cut}}$ (\footnotesize{kpc}) &$\sigma$ (\footnotesize{km\,s$^{-1}$})\\
\noalign{\smallskip}
\hline
\hline
\noalign{\smallskip}
\noalign{\smallskip}
\#1 &  $-1.95^{+0.10}_{-0.19}$  & 0.17 $^{+0.15}_{-0.22}$  &  0.58 $\pm$0.01  &  30.58$^{+0.35}_{-0.51}$  & 112.9$^{+3.6}_{-2.1}$  & [1000] & 1015$^{+7}_{-6}$ \\
\noalign{\smallskip}
\hline
\noalign{\smallskip}
\#2 &  -28.02$^{+0.26}_{-0.17}$ & -36.02$^{+0.27}_{-0.21}$ & 0.70$\pm$0.02 & 39.02$^{+2.23}_{-1.69}$ & 16.5$^{+2.7}_{-3.9}$  & [1000] & 331$^{+13}_{-9}$\\
\noalign{\smallskip}
\hline
\noalign{\smallskip}
\#3 & $-48.65^{+0.13}_{-0.49}$  & $-51.35^{+0.30}_{-0.22}$  &  0.35 $\pm$0.02  &  126.48$^{+7.11}_{-4.42}$  &  64.2$^{+6.8}_{-9.6}$  & [1000] & 286$^{+24}_{-16}$ \\
\noalign{\smallskip}
\hline
\noalign{\smallskip}
\#4 &  $17.62^{+0.28}_{-0.18}$  & 46.90 $^{+0.36}_{-0.28}$  &  0.15 $\pm$0.02  & 54.66$^{+3.51}_{-4.83}$  &  110.5$^{+1.2}_{-2.1}$  & [1000] & 688$^{+9}_{-17}$ \\
\noalign{\smallskip}
\hline
\noalign{\smallskip}
\#5 &  $-17.22^{+0.17}_{-0.18}$  & 101.85 $^{+0.08}_{-0.07}$  &  0.44 $\pm$0.05  &  62.29$^{+5.14}_{-4.61}$  &  2.1$^{+0.5}_{-0.1}$  & [1000] & 263$^{+8}_{-7}$ \\
\noalign{\smallskip}
\hline
\noalign{\smallskip}
\#6 &  [0.0]  & [0.0]  &  [0.2]  &  [34.0]  &  3.95$^{+0.57}_{-0.89}$  & 92.08$^{+6.50}_{-7.91}$ & 284$\pm$8 \\
\noalign{\smallskip}
\hline
\noalign{\smallskip}
\#7 &  [3.16]  & [-11.10]  &  0.22 $\pm$0.02  &  103.56$^{+7.09}_{-7.95}$  &  [0.15]  & 43.17$^{+1.34}_{-1.02}$ & 152$^{+2}_{-1}$ \\
\hline
\noalign{\smallskip}
L$^*$ elliptical galaxy & --  & --  & -- & -- & [0.15] & 52.48$^{+2.17}_{-0.89}$ & 148$^{+2}_{-3}$ \\
\noalign{\smallskip}
\hline
\hline
\end{tabular}
\caption{PIEMD parameters inferred for the five dark matter clumps considered in the optimization procedure. Clumps \#6 and \#7 are galaxy-scale halos that were modeled separately from scaling relations, to respectively model the BCG of the cluster as well as the cluster member responsible for the four multiple-images of \textit{SN Refsdal}.
Coordinates are given in arcseconds with respect to
$\alpha=177.3987300, \delta=22.3985290$.
The ellipticity $e$ is the one of the mass distribution, in units of $(a^2+b^2)/(a^2-b^2)$.
The position angle $\theta$ is given in degrees and is defined as the direction of the semi-major axis of the iso-potential, counted counterclockwise from the horizontal axis (being the R.A. axis).
Error bars correspond to the $1\sigma$ confidence level.
Parameters in brackets are not optimized. 
Concerning the scaling relations, the reference magnitude is $mag_{\rm F814W} = 20.65$.
}
\label{SLmodel_res}
\end{center}
\end{table*}


\subsection{Methodology}
Since our method to create the MACSJ1149 mass model closely follows the one used by \cite{jauzac14,jauzac15b}, we here only give a brief summary and refer the reader to \cite{kneib96,smith05,verdugo11,richard11b} for more details.
Our mass model combines large-scale dark-matter haloes to model the cluster-scale mass components  and galaxy-scale haloes to model individual cluster members, typically large elliptical galaxies or galaxies that affect strong-lensing features due to their proximity to multiple images.
As in our previous work, all mass components are modeled as Pseudo Isothermal Elliptical Mass Distribution \citep[PIEMD,]{limousin05,jauzac14,jauzac15b}, characterized by a velocity dispersion $\sigma$, a core radius $r_{core}$, and a cut radius $r_{cut}$.
In this parametric approach, haloes are not allowed to contain zero mass, and hence the relative probability of different models meeting the observational constraints is quantified by the $\chi^{2}$ and $RMS$ statistics.

For the PIEMD models added to parameterize cluster members (mass perturbers) we fix the PIEMD parameters (center, ellipticity and position angle) at the values measured from the cluster light distribution \citep[see, e.g.][]{kneib96,limousin07b,richard10} and assume empirical scaling relations to relate the dynamical PIEMD parameters (velocity dispersion and cut radius) to the galaxies' observed luminosity \citep{richard14}.
For an $L^{\ast}$ galaxy, we optimize the velocity dispersion between 100 and 250 km s$^{-1}$ and force the cut radius to less than 70 kpc to account for tidal stripping of galactic dark-matter haloes \citep{limousin07b,limousin09a,natarajan09,wetzel10}.

Finally, one more parameter that is fixed, but plays an important role in the $\chi^2$ computation, is the positional uncertainty of the multiple images. Indeed, it will affect the derivation of errors, i.e. a smaller positional uncertainty will generally result in a smaller statistical uncertainty, thus leading to an underestimation of the statistical error budget.

With HFF-like data, the average astrometric precision of the image position is of the order of 0.05$\arcsec$. However, parametric cluster lens models often fail to reproduce or predict image positions to this precision, yielding instead typical image-plane RMS values between $0.2\arcsec$ and a few arc seconds. In this work, we use a positional uncertainty of $0.5\arcsec$.

\subsection{Results}
Our mass model includes five cluster-scale haloes, referred to as \#1, \#2, \#3, \#4 and \#5 in Table~\ref{SLmodel_res}. An additional cluster-scale halo, South East of the BCG, was requested by the model compare to the pre-HFF mass model presented in \cite{richard14}. Indeed, its necessity was confirmed by both a smaller RMS value and a better reduced $\chi^{2}$ compare to a four cluster-scale halos mass model.
During the optimisation process, the position of these large-scale halos is allowed to vary within 20$\arcsec$ of the associated light peak. In addition, we limit the ellipticity, defined as $e=(a^2+b^2)/(a^2-b^2)$, to values below 0.7,  while the core radius and the velocity dispersion are allowed to vary between 1$\arcsec$ and 35$\arcsec$, and 100 and 2\,000 km\,s$^{-1}$, respectively. 
The cut radius, by contrast, is fixed at 1\,000\,kpc, since strong-lensing data alone do not probe the mass distribution on such large scales. 
In addition to these five cluster-scale dark-matter halos, we include galaxy-scale perturbations induced by 216 probable cluster members \citep[][]{richard14} by assigning a galaxy-scale halo to each of them. Finally, we add two galaxy-scale halos to model the BCG of the cluster (Clump \#6 in Table~\ref{SLmodel_res}), as well as the cluster member lensing \textit{SN Refsdal} (Clump \#7 in Table~\ref{SLmodel_res}). 
Using the set of the most securely identified multiply imaged galaxies described in Sect.~\ref{Mul_SL} and shown in Fig.~\ref{multiples}, we optimise the free parameters of this mass model using the publicly available \textsc{Lenstool} software\footnote{http://projects.lam.fr/repos/lenstool/wiki}.

The best-fit model optimised in the \emph{image plane} predicts image positions that agree with the observed positions to within an RMS of 0.91\arcsec, a value that is slightly higher than the one published in \cite{richard14}. 
This increase is in part caused by the high individual $\chi^{2}$ value of image 3.3 ($\chi^{2}=13$), which we nonetheless consider a robust identification (see discussion in Sect.\ref{Mul_SL}). 

The parameters describing our best-fit mass model are listed in Table~\ref{SLmodel_res}. Although allowed to vary within 20$\arcsec$ of their associated light peak, the final positions of the five cluster-scale halos predicted by the model coincide much more closely with the light peaks. 
In order to integrate the mass map within annuli, we choose the location of the overall BCG, i.e., $\alpha=177.3987300, \delta=22.3985290$ degrees, as the cluster centre. 
The two-dimensional (cylindrical) mass within 80$\arcsec$ is then $M({R}{<}500~$kpc$) = (6.29\pm 0.03) \times 10^{14}$\,M$_{\odot}$, slightly lower than the previous result by \cite{smith09} of $M(R{<}500~$kpc$) = (6.7\pm 0.4) \times 10^{14}$\,M$_{\odot}$, but within their error bars. The mass of $(1.71\pm 0.20) \times 10^{14}$\,M$_{\odot}$ within a 30$\arcsec$ ($\sim$200~kpc) aperture reported by \cite{zitrin11} agrees with the value from our new mass model of $M(R{<}200~$kpc$) = (1.840\pm 0.006) \times 10^{14}$\,M$_{\odot}$.

\section{Discussion}
\label{discussion}

\subsection{The lensing power of MACSJ1149}
\label{lenspower}
\begin{figure}
\begin{center}
\includegraphics[width=0.5\textwidth]{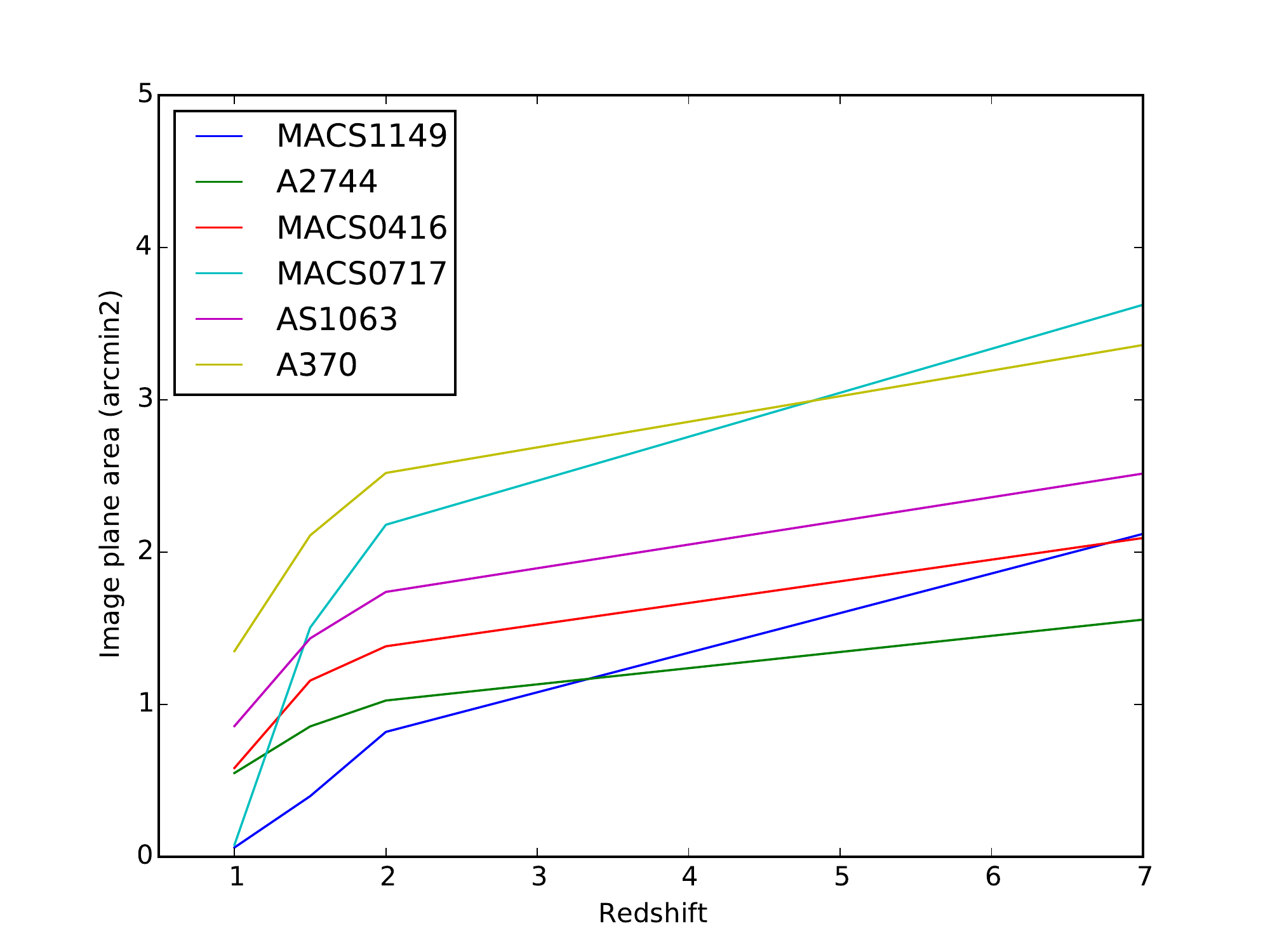}
\caption{Evolution of the surface area in the image plane within which multiple images are observed, as a function of the 
source redshift. Compared to the five other Frontier Fields clusters, MACSJ1149 features a much smaller multiple-image 
 region at low redshifts.}
\label{surf}
\end{center}
\end{figure}

\begin{table*}
\begin{center}
\begin{tabular}[h!]{ccccccc}
\hline
\hline
\noalign{\smallskip}
Component  & S1 & S2 & S3 & S4 & SX & SY \\
\hline
\hline
R.A. & 177.39823 & 177.39771 & 177.39737 & 177.39780 & 177.40024 & 177.4038 \\
Dec. & 22.395628 & 22.395789 & 22.395542 & 22.395172 & 22.396811 & 22.402149 \\
$\mu$ & 22.4$\pm$2.0 & 18.9$\pm$2.3 & 19.7$\pm$1.7 & 9.2$\pm$0.8 & 4.3$\pm$0.1 & 3.6$\pm$0.1 \\
\hline
 $\Delta$t$_{CATS}$ & 0.0 & 90$\pm$17 & 30$\pm$35 & -60$\pm$41 & 449$\pm$45 & -4654$\pm$358 \\
 $\Delta$t$_{CATS-src}$ & 0.0 & -0.8$\pm$1.6 & 8.1$\pm$1.6 & 0.2$\pm$0.4 & 361$\pm$42 & -5332$\pm$357 \\
 \hline
$\Delta$t$_{Sharon+15}$ & 0.0 & 2.0 & -5.0 & 7.0 & 237$^{+37}_{-50}$ & -4251$^{+369}_{-373}$ \\
 $\Delta$t$_{Oguri+15}$ & 0.0 & 9.2 & 5.2 & 22.5 & 357.1 & -6193.5 \\
 $\Delta$t$_{Diego+15}$ & -- & -- & -- & -- & 376$\pm$25 & -3325$\pm$763 \\
 \hline
$\Delta$t$_{Die-a}$ & 0.0 & -17$\pm$19 & -4.0$\pm$27 & 74$\pm$43 & 262$\pm$55 & -4521$\pm$524 \\
$\Delta$t$_{Gri-g}$ & 0.0 & 10.6$^{+6.2}_{-3.0}$ & 4.8$^{+3.2}_{-1.8}$ & 25.9$^{+8.1}_{-4.3}$ & 361$^{+19}_{-27}$ & -6183$^{+160}_{-145}$\\
$\Delta$t$_{Ogu-g}$ & 0.0 & 8.7$\pm$ 0.7 & 5.1$\pm$0.5 & 18.8$\pm$1.7 & 311$\pm$24 & -5982$\pm$287 \\
$\Delta$t$_{Ogu-a}$ & 0.0 & 9.4$\pm$1.1 & 5.6$\pm$0.5 & 20.9$\pm$2.0 & 336$\pm$21 & -6239$\pm$224 \\
$\Delta$t$_{Sha-g}$ & 0.0 & 6$^{+6}_{-5}$ & -1$^{+7}_{-5}$ & 12$^{+3}_{-3}$ & 277$^{+11}_{-21}$ & -5016$^{+281}_{-15}$\\
$\Delta$t$_{Sha-a}$ & 0.0 & 8$^{+7}_{-5}$ & 5$^{+10}_{-7}$ & 17$^{+6}_{-5}$ & 233$^{+46}_{-13}$ & -4860$^{+126}_{-305}$\\
$\Delta$t$_{Zit-g}$ & 0.0 & -161$\pm$97 & -149$\pm$113 & 82$\pm$51 & 224$\pm$262 & -7665$\pm$730\\
 \hline
\end{tabular}
\caption{Time delays obtained for each image of \textit{SN Refsdal}, given in days. 
The first part of the Table gives the coordinates of the SN images used in this analysis, with the ones for SX and SY predicted by our best-fit mass model. We also quote the magnification predicted by our best-fit mass model, $\mu$.
The second part of the Table presents the time delays we measured with our analysis, following the two different methods presented in Sect.~\ref{discussion}, $\Delta$t$_{CATS}$ and $\Delta$t$_{CATS-src}$.
We then give the results obtained by pre-HFF analysis in the third portion of the Table: \citet{sharon15}, $\Delta$t$_{Sharon+15}$, \citet{oguri15}, $\Delta$t$_{Oguri+15}$, and \citet{diego15}, $\Delta$t$_{Diego+15}$. 
And finally the fourth and final section of the table lists the results obtained with the most recent HFF analysis presented by \citet{treu15}: Diego's model, $\Delta$t$_{Die-a}$, \citet{grillo15} model, $\Delta$t$_{Gri-g}$, \citet{kawamata15} models, $\Delta$t$_{Ogu-g}$ and $\Delta$t$_{Ogu-a}$, Sharon's models, $\Delta$t$_{Sha-g}$ and $\Delta$t$_{Sha-a}$, and Zitrin's model, $\Delta$t$_{Zit-g}$.
for comparison with our analysis, $\Delta$t$_{CATS}$ and $\Delta$t$_{CATS-src}$.
}
\label{sn_timedelays}
\end{center}
\end{table*}

The small number of 22 multiple-imaged systems identified in MACSJ1149 from HFF data is surprising, compared to the 34 and 51 systems found in equally deep HST images of the first two HFF clusters, Abell 2744 and MACSJ0416. Because of this discrepancy, we carefully examined the MUSE data for several sources at $z\sim1$ within the central arcmin$^2$ of the cluster core, but confirmed all of them to be single images, as predicted. 

In an attempt to find the root cause of the relatively modest lensing power of MACSJ1149, we compute the surface area in the image plane within which we expect multiple images (Fig.~\ref{surf}). For all six Frontier Field clusters we find this area to increase with source redshift, starting at an average value of $\sim$1 arcmin$^2$ at $z=1$ and reaching values around 2.5 arcmin$^2$ at $z=7$, although with large cluster-to-cluster differences. 

Whereas the overall shape of these curves is very similar for four of the six HFF clusters, to the extent that they are essentially scaled versions of each other, two systems exhibit a different trend with redshift. For both MACSJ1149 and MACSJ0717, the curves in Fig.~\ref{surf} start at a very small surface area at $z=1$ ($\sim0.1$ arcmin$^2$) but then rise much more steeply than those of the remaining HFF targets. 
Part of this effect is due to their being at higher redshift ($z=0.55$), causing their corresponding critical surface mass density to decrease more rapidly with increasing source redshift. We argue that the trend is also rooted in both of these clusters sharing a similar morphology
or, more generally, in the fact that extremely disturbed clusters (both MACSJ0717 and MACSJ1149 consist of more than three subclusters) are at a disadvantage for lensing background sources at low redshift, owing to disjoint critical lines, but gain on less complex systems at higher source redshifts at which the critical lines for strong lensing join. 
Being the much more massive system, MACSJ0717 features a larger image-plane area throughout though, thereby ``catching up" much faster than the less massive MACSJ1149.

In conclusion, we trace the small number of multiple images in MACSJ1149 to the system's complex and extended morphology which leads to the smallest area for multiple imaging of sources at redshifts $z<3$ among all \textit{Hubble Frontier Fields} target clusters. Following the same argument, we expect the two final HFF targets, Abell 370 and AS1063, to produce much larger number of multiple images when they are observed with HST in the coming months.


\begin{figure*}
\begin{center}
\includegraphics[width=0.33\textwidth]{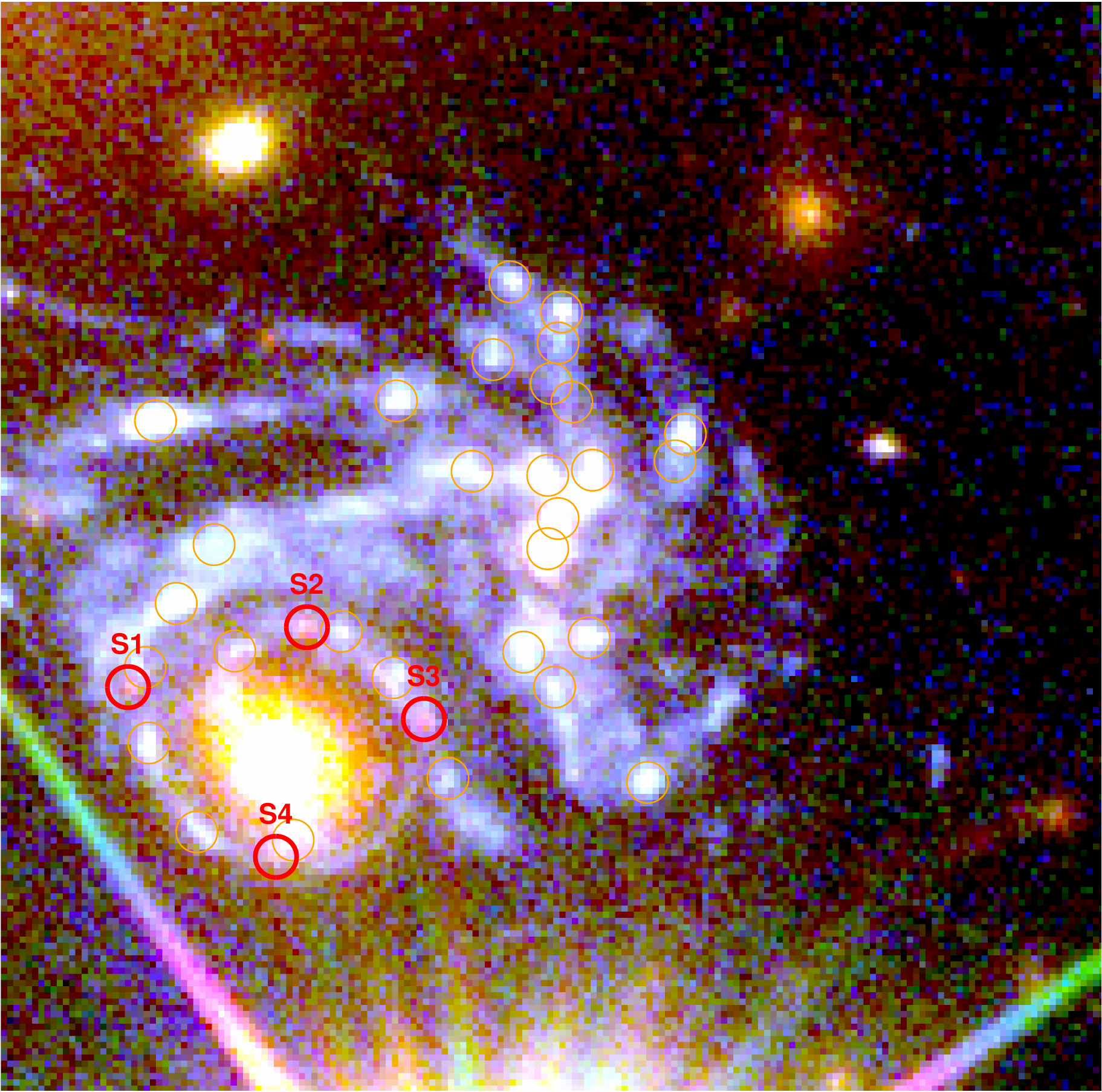}
\includegraphics[width=0.33\textwidth]{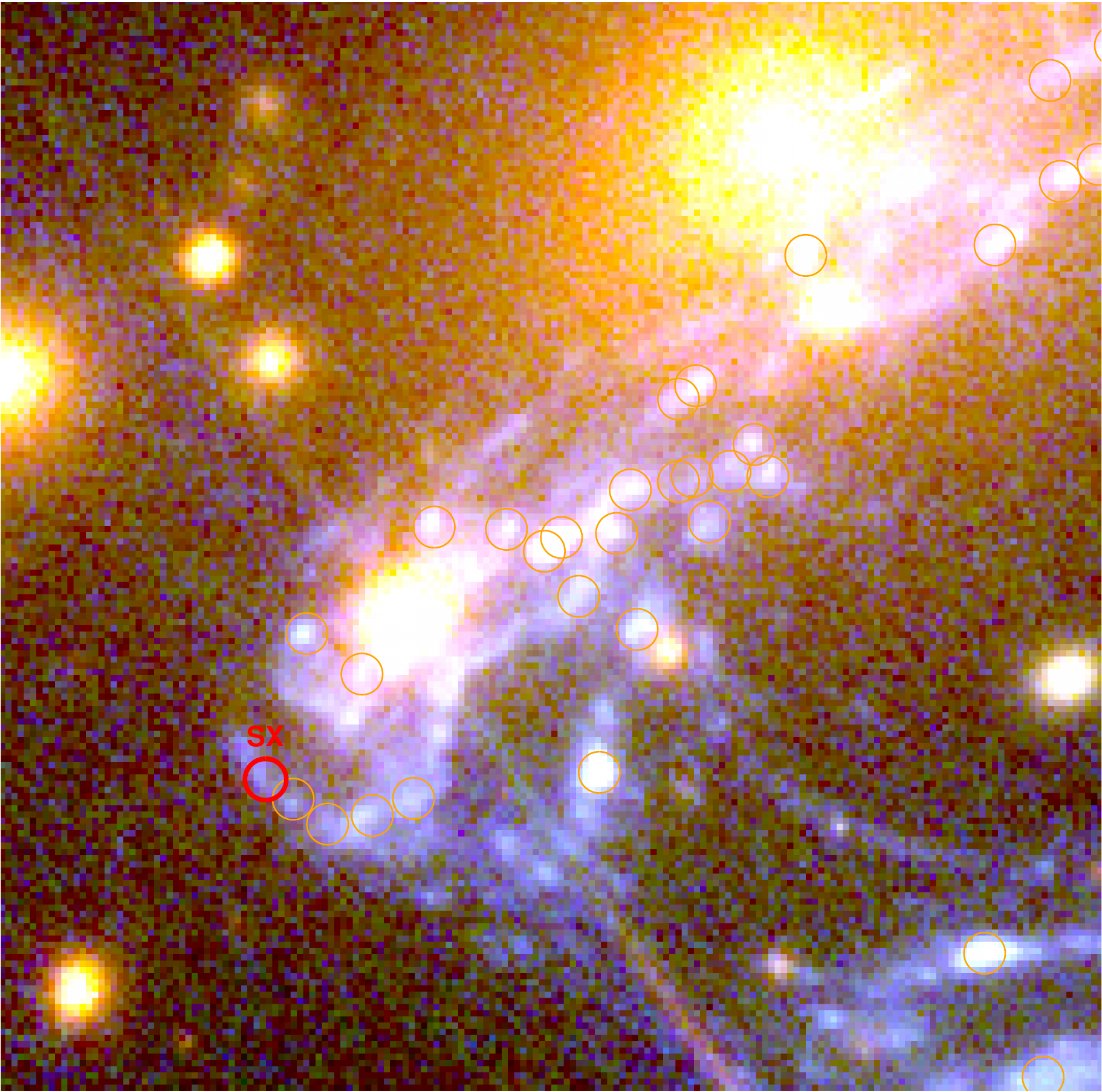}
\includegraphics[width=0.33\textwidth]{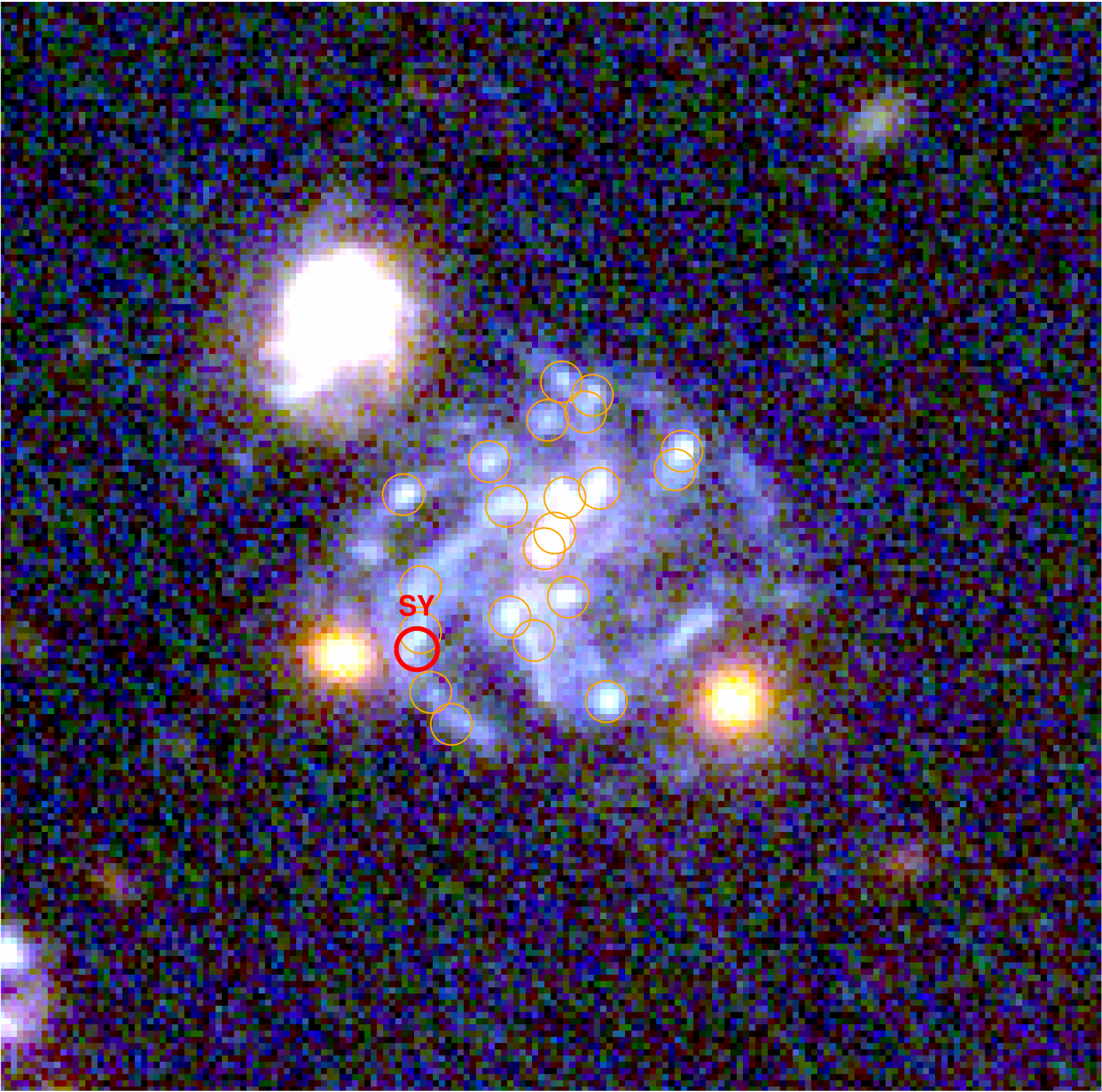}\\
\caption{Zoom on System \#1, with orange circles highlighting the 24 components of the spiral used for our mass model. \textit{Left panel:} Image 1.1 with the Einstein Cross formed by images S1, S2, S3, and S4 of \textit{SN Refsdal}. \textit{Middle panel:} Image 1.2 with the predicted location of \textit{SN Refsdal} in red, SX, predicted to appear $\sim$1.8 years after S1, i.e. in January 2016. \textit{Right panel:} Image 1.3 with the predicted position of \textit{SN Refsdal}, SY, where it would have been visible $\sim$ 11.5 years ago following the predictions from our model.}
\label{sn_zoom}
\end{center}
\end{figure*}

\subsection{\textit{SN Refsdal}}
\subsubsection{Different studies of SN Refsdal}

Since the discovery of \textit{SN Refsdal}, three new pre-HFF strong-lensing analyses have attempted to precisely compute the time delays between the multiple images of the SN \citep[][]{sharon15,oguri15,diego15}, using both parametric and free-form approaches. 
While this paper was under review, three additional strong-lensing groups have presented predictions using both HFF, VLT, and Keck data \citep[][]{treu15,grillo15,kawamata15}\footnote{Another recent analysis focuses on the environment provided the host of \emph{SN Refsdal} \citep[][]{karman15}, a subject that is outside the scope of our paper.}.
All the models predict six multiple images of \emph{SN Refsdal} in total: one in each of the three multiple images of the face-on spiral (1.1, 1.2. 1.3), with the one in image 1.1 quadruply imaged by a cluster member, resulting in four separate components named S1, S2, S3 and S4, as shown in Fig.~\ref{sn_zoom}.

\paragraph*{Sharon \& Jonhson (2015), Diego et al. (2015) \& Oguri (2015):} Using \textsc{Lenstool}, \cite{sharon15} investigated different models and found the order of appearance for the four images within image 1.1 to be S3-S1-S2-S4. As for the remaining two images of SN Refsdal, \cite{sharon15}  claim that the first one, SY, appeared in image 1.3 $\sim$12~years ago, while the sixth image, SX, is predicted to appear in image 1.2 in September 2015. By contrast, \cite{oguri15}, using the public code \emph{glafic} \citep[][]{oguri10}, arrives at a different order of appearance of the SN within image 1.1, namely S1-S3-S2-S4 (see Table~\ref{sn_timedelays}), claims that SY appeared $\sim$17~years ago, and predicts that SX is to appear in November 2015. Finally, \cite{diego15}, using a free-form mass modeling technique, WSLAP+ \citep[][]{diego05,diego07,sendra14}, do not address the order of appearance of images S1, S2, S3 and S4, but conclude that SY appeared $\sim$9 years ago and predict SX to appear between November 2015 and January 2016.

\paragraph*{Treu et al. (2015), Kawamata et al. (2015) \& Grillo et al. (2015):} More recently, \cite{treu15} presented predictions based on seven different mass models. These new measurements were made using HFF mass models, and taking advantage of new spectroscopy coming from HST with GLASS data (PI: Treu), as well as the complete MUSE DDT program 294.A-5032(A) (PI: Grillo; for our model we only had access to 4 hours of the data as detailed in Sect.~\ref{observations}).
The teams involved in this analysis (Sharon, Zitrin, Diego, Grillo and Oguri) identified independently new multiple image systems and then voted to select the most secured identifications following the same criteria as presented in \cite{wang15}.
While our two analyses were done separately, we find a good agreement in terms of new multiple-image systems: system \#22, system \#29, system \#21, system \#15, system \#16, system \#31, system \#17, system \#18 and system \#34 correspond respectively to system \#110, system \#21, system \#24, system \#26, system \#27, system \#28, system \#203, system \#204, and system \#205 in \cite{treu15}.
There is no clear discrepancy between our spectroscopic measurements and the ones presented in \cite{treu15} and \cite{grillo15}. However, looking at the cluster members, \cite{treu15} present 170 spectroscopic identifications while we only identify 57 (see Sect.~\ref{observations}).
\cite{grillo15} independently presented their strong-lensing mass model, as well as the complete dataset they obtained with the MUSE program 294.A-5032(A). \cite{kawamata15} also presented the Oguri's models in their paper. The results did not change from the ones presented in \cite{treu15}.

The five teams involved in the \cite{treu15} \textit{SN Refsdal} analysis predicted time delays for images S1, S2, S3 and S4, with different order of appearance and occurring within 12 and up to 240 days, depending on the model. The appearance of SX at peak brightness is predicted during the first half of 2016, while SY would have appeared between 1994 and 2004. However, magnification estimates from the different models suggest that it would have been too faint to be observable from archival data.

We refer the reader to Table~\ref{sn_timedelays} for the exact values of the time delays predicted by these eight different measurements, as well as the ones found by our analysis. For consistency, we use in all cases S1 as our reference and adopt the established naming convention (SX and SY) for  the two other images of \textit{SN Refsdal} in image 1.2 and 1.3, respectively. We note that some of the teams in \cite{treu15} had two sets of predictions, as they run different mass models using two different sets of multiple images (`gold', the most secured ones, and `all', for the gold images plus a subset of less secured identifications). In Table~\ref{sn_timedelays} we keep the same notation as in \cite{treu15} to limit confusion.

\subsubsection{Predictions for S1, S2, S3, and S4}
Our modelling efforts do not use time delays as constraints but, as explained in Sect.~\ref{Mul_SL}, we include the position of the observed four images of \textit{SN Refsdal}, as well as an independent PIEMD to model the cluster member responsible for the lensing of the spiral arm of image 1.1, and hence for the formation of the Einstein cross images. 
The delays between the appearance of the four images S1 to S4 depend on the slope of the mass profile at the location of the lensing galaxy. The current uncertainties in the delays computed by various teams are caused by the fact that this slope can ultimately only be constrained by the highly unlikely detection of the central, fifth image of the SN, which is demagnified by the galaxy lens. 


The arrival-time surface for a light ray emitted by a source, at the location $\vec{\beta}$, traversing the cluster at the location $\vec{\theta}$, is written as \citep[e.g., ][]{bible2} :
\begin{equation}
\tau(\vec{\theta},\vec{\beta}) = \frac{1 + z_{cl}}{c} \frac{D_{OL} D_{OS}}{D_{LS}} \left[\frac{1}{2} (\vec{\theta} - \vec{\beta})^{2} - \phi(\vec{\theta}) \right]
\label{timedel_eq}
\end{equation}
where $\vec{\theta}$ and $\vec{\beta}$ are the image and the source plane positions respectively, $z_{cl}$ is the redshift of the cluster, $D_{OL}$, $D_{OS}$ and $D_{LS}$ are the cosmological distances between the observer and the lens, the observer and the source, and the lens and the source respectively. Finally, $\phi(\vec{\theta})$ represents the gravitational potential of the cluster.
The measurement of the time delays is thus highly sensitive to the source plane position.

\textsc{Lenstool} measures time delays following this equation. However, for the different multiple images of a system, the model will predict a slightly different position of the source.\textsc{Lenstool} time delay-measurement for one system will thus be based on a slightly different source position for each image, thus giving different departure times. The entries $\Delta t_{CATS}$ in Table~\ref{sn_timedelays} correspond to these measurements.

To overcome this issue we repeated the measurements following Eq.~\ref{timedel_eq} but using an analytical method. Indeed, while the image positions are taken as the observed ones (or model predicted ones for SX and SY), we set $\vec{\beta}$ to the barycenter of the predicted source positions for the four SN images, in order to establish the same departure location and time in all measurements. Finally, the gravitational potential is derived using \textsc{Lenstool}, and then inserted in the equation. This `analytical' way of measuring arrival-time surfaces (as opposed to the numerical approach taken in our previsou use of \textsc{Lenstool}) has proven to be much more accurate. The entries $\Delta t_{CATS-src}$ in Table~\ref{sn_timedelays} correspond to these measurements.

Depending on the method used to measure the time delays, our best-fit mass model leads to an order of appearance in image 1.1 of S4-S1-S3-S2, and S2-S1-S4-S3 following either the \textsc{Lenstool} implementation or the analytical method respectively.
The \textsc{Lenstool} method predicts considerably longer delays of $\Delta t_{S2-S1} = 90\pm 17$~days, $\Delta t_{S3-S1} = 30\pm 35$~days, and $\Delta t_{S4-S1} = -60\pm 41$~days, compared to $\Delta t_{S2-S1} = -0.6\pm$1.6~days, $\Delta t_{S3-S1} = 8.1\pm 1.6$~days, and $\Delta t_{S4-S1} = 0.2\pm 0.4$~days obtained with the `analytical' version. These last measurements are much more realistic, considering the measurements obtained through the detailed study of the SN itself by \cite{kelly15} \citep[also presented in ][]{treu15}. 
Our best-fit analysis concludes that the four images appeared within $\sim$5 months following our first measurements, and within $\sim$10 days following the analytical measurements (see Table~\ref{sn_timedelays}, $\Delta$t$_{CATS}$ and $\Delta$t$_{CATS-src}$ respectively). The error bars are derived while measuring the time surfaces for each MCMC realizations.

Most of the models \citep[][]{oguri15,sharon15,treu15,grillo15,kawamata15} claim different orders of appearance for images S1--S4, which is puzzling in particular with regard to \cite{sharon15} and HFF Sharon's model presented in \cite{treu15} who use the same software as us, \textsc{Lenstool}. 

There are important differences between their analysis and ours. The first one lies in the number of multiple images used, as \cite{sharon15} work with a pre-HFF mass model and the list of multiple images published in \cite{johnson14}, and \cite{treu15} use a different set of HFF multiple images. 
We initially thought that an important difference between our analysis and the one by \cite{sharon15} resided in a mis-identified spectroscopic redshift of $z{=}2.497$ for system \#3, as published in \cite{smith09}, whereas we revised this value to  $z{=}3.128$ using the recent MUSE observation of this system \citep[also confirmed by][]{treu15,grillo15}. Running the pre-HFF strong-lensing model of \cite{richard14} with either redshift of system \#3, we find the erroneous earlier redshift to result in time delays for images S1-S4 that are much smaller than those obtained with the new redshifts and either the pre-HFF or our latest lens model. This trend is observed using both time surface's measurement methods. We thus concluded that this mis-identified redshift had a major impact on the time delay values and could be responsible for the bulk of the discrepancy between our findings and those of \cite{sharon15}. However, while this paper was under review, \cite{treu15} presented a revised version of Sharon's model using the correct redshift for system \#3, and their results did not change.

The large differences between all the recent models' estimations still remain puzzling. We thus investigated other sources of error, and identified another systematic effect that has a significant impact on the values of the time delays for S1--S4: the value of $r_{\rm core}$ that characterises the shape of gravitational potential of the elliptical galaxy lensing \textit{SN Refsdal} in image 1.1 (Clump \#7 in Table~\ref{SLmodel_res}). The choice of  r$_{\rm core}$ affects both the order of appearance of the four images and the associated time delays while using the \textsc{Lenstool} implementation. Indeed, we estimate this effect to produce changes in $\Delta t_{Si-S1}$ ($i{=}2, 3,4 $) of up to 30 days. Moreover, while the `analytical' method does not show any differences in time delay values, it however changes the order of appearance of the different images S1-S4.
The best-fit parameters obtained for this galaxy are listed in Table~\ref{SLmodel_res} (Clump \#7). However, they should be taken with caution as explained above.

Although the same sources of systematic errors were identified using both methods, we revise our conclusions, and link the discrepancy between our first measurements and other independent analysis to the incorrect method that is implemented in \textsc{Lenstool} at the moment. We thus consider our second measurements, $\Delta$t$_{CATS-src}$, more accurate. A modification of the algorithm will be implemented according to the conclusions of this paper, and will be made public early 2016.

\subsubsection{Predictions for SX and SY}
All strong-lensing analyses conducted so far (including ours) agree that the SN image SY was the first one to appear in image 1.3, from our analysis $\sim$15~years before S1, i.e., $\Delta t_{\rm SY-S1} = (-5332\pm 357)$ days (and $\sim$13~years ago, $\Delta t_{\rm SY-S1} = (-4654\pm 358)$ days, following the \textsc{Lenstool} prediction).
With our first measurement, the appearance of image SX in image 1.2 is expected ${\sim}1$ years after S1 ($\Delta t_{\rm SX-S1} = 449\pm 45$ days) and $\sim$12~months after S2, i.e., in November 2015, assuming that S2 appeared in November 2014 \citep[][]{kelly15}. 
Our analytical measurement predicts an appearance of SX $\sim$1~year after S1 (with an appearance of S1, S2, S3 and S4 with 8 days), in November 2015 plus/minus 1 month, and thus also agreeing with the \textsc{Lenstool} prediction.

We can also compare our results regarding the time delays between SX and SY with those of \cite{sharon15}, \cite{oguri15} and \cite{diego15}. 
Our initial measurement predicts $\Delta t_{\rm SX-SY} = (5103\pm 361)$ days, $\sim$14~years, which is in good agreement (within the uncertainty of $\sim$1 year) with the value of $\Delta t_{\rm SX-SY} = 4488$~days, $\sim$12.5 years, reported by \cite{sharon15}, and close to the value found by \cite{diego15} of $\Delta t_{\rm SX-SY} = 3701$~days, $\sim$10 years (with an uncertainty of $\sim$2 years).  Our analytical value of $\Delta t_{\rm SX-SY} = (5693\pm 361)$ days, $\sim$16~years, remains in really good agreement with the \textsc{Lensool} measurement.
All of these values are smaller than that of $\Delta t_{\rm SX-SY} = 6650.6$~days, $\sim$18 years, published in \cite{oguri15}. 

To summarise, considering the statistical and systematic errors, we conclude that image SX of \textit{SN Refsdal} should appear between November 2015 and January 2016, in good agreement with predictions from \cite{diego15} and \cite{grillo15}.

Although all the models discussed here make different predictions for the appearance of image SX,  they agree that it will occur in the sufficiently near future to render the event observable with a reasonable investment of time and resources. However, ultimately only photometric follow-up observations of this source will reveal the true time delays. Once obtained, such measurements can be included as constraints in the mass models, enabling us to determine to unprecedented precision the parameters of the lensing galaxy (Clump \#7), and to study the stellar and dark matter distribution within it, thereby also adding greatly to our efforts to reveal the nature of dark matter \citep{massey15}.

While this manuscript of this paper was under review, image SX of \textit{SN Refsdal} was detected in HST/WFC3 images inobservations taken in December 2015, making our predictions correct \citep[][]{kelly15b}, in excellent agreement with our predictions.

\section*{Acknowledgments}
The authors thank Prof. Keren Sharon for fruitful discussions, and important suggestions. MJ thank Prof. Tommaso Treu for his comments and suggestions.
This work was supported by the Science and Technology Facilities Council [grant number ST/L00075X/1 \& ST/F001166/1] and used the DiRAC Data Centric system at Durham University, operated by the Institute for Computational Cosmology on behalf of the STFC DiRAC HPC Facility (www.dirac.ac.uk [www.dirac.ac.uk]). This equipment was funded by BIS National E-infrastructure capital grant ST/K00042X/1, STFC capital grant ST/H008519/1, and STFC DiRAC Operations grant ST/K003267/1 and Durham University. DiRAC is part of the National E-Infrastructure. 
JR acknowledges support from the ERC starting grant CALENDS and the CIG grant 294074. 
MJ, EJ, and ML acknowledge the M\'esocentre d'Aix-Marseille Universit\'e (project number: 15b030).  This study also benefited from the facilities offered by CeSAM (CEntre de donn\'eeS Astrophysique de Marseille ({\tt http://lam.oamp.fr/cesam/}). MJ thanks K. Sharon for fruitful discussions. ML acknowledges the Centre National de la Recherche Scientifique (CNRS) for its support.
EJ and ML acknowledge the Centre National d'Etude Spatial (CNES) for its support. 
KK acknowledges postgraduate funding from the NRF/SKA SA Project.
JPK acknowledges support from the ERC advanced grant LIDA. PN acknowledges support from the National Science Foundation via the grant AST-1044455, AST-1044455, and a theory grant from the Space Telescope Science Institute HST-AR-12144.01-A. RM is supported by the Royal Society.
Based on observations made with the NASA/ESA Hubble Space Telescope, obtained from the data archive at the Space Telescope Science Institute. STScI is operated by the Association of Universities for Research in Astronomy, Inc. under NASA contract NAS 5-26555. 
Based on observations made with the European Southern Observatory Very Large Telescope (ESO/VLT) at Cerro Paranal, under programme ID 294.A-5032 (PI: Grillo).
Based on observations obtained at the Gemini Observatory, which is operated by the Association of Universities for Research in Astronomy, Inc., under a cooperative agreement with the NSF on behalf of the Gemini partnership (as at the date that the data described in this article were obtained): the National Science Foundation (United States), the Science and Technology Facilities Council (United Kingdom), the National Research Council (Canada), CONICYT (Chile), the Australian Research Council (Australia), Minist\'{e}rio da Ci\^{e}ncia, Tecnologia e Inova\c{c}\~{a}o (Brazil) and Ministerio de Ciencia, Tecnolog\'{i}a e Innovaci\'{o}n Productiva (Argentina).
MJ thanks the Department of Astronomy at Yale University for their hospitality.

\bibliographystyle{mn2e}
\bibliography{reference}


\newpage
\begin {table}
\begin{tabular}{cc}
\, & \,\\
\end{tabular}
\caption{ Multiply imaged systems considered in this work.
$^{+}$ Thanks to the VLT/MUSE data, we were able to revise spectroscopic redshift of system \#1, from $z=1.491$ as in \citet{smith09} to $z=1.4888$.
$^{\ast}$ indicate the different components of system \#1 we have used for our model, following the decomposition presented in \citet{rau14}.
We include the predicted magnification given by our model. Some of the magnitudes are not quoted because we were facing deblending issues that did not allow us to get reliable measurements. The flux magnification factors come from our best-fit mass model, with errors derived from MCMC sampling.
}
\label{multipletable}
\end{table}

\scriptsize

\begin{center}
\par
\tablefirsthead{\hline          \multicolumn{1}{c}{\textbf{ID}} &
                                \multicolumn{1}{c}{\textbf{R.A.}} &
                                \multicolumn{1}{c}{\textbf{Decl.}} &
                                \multicolumn{1}{c}{\textbf{$z$}} &
                                \multicolumn{1}{c}{m$_{F814W}$} &
                                \multicolumn{1}{c}{$\mu$}
                                \\ \hline }

\tablehead{\hline \multicolumn{5}{l}{\small\sl continued from previous page}\\
                         \hline \multicolumn{1}{c}{\textbf{ID}} &
                                \multicolumn{1}{c}{\textbf{R.A.}} &
                                \multicolumn{1}{c}{\textbf{Decl.}} &
                                \multicolumn{1}{c}{\textbf{$z_{\rm model}$}} &
                                \multicolumn{1}{c}{m$_{F814W}$} &
                                \multicolumn{1}{c}{$\mu$}
                                \\ \hline  }
\tabletail{\hline\multicolumn{5}{r}{\small\sl continued on next page}\\\hline}
\tablelasttail{\hline}
\par
\begin{supertabular}{llllll}
\hline 
  $^{+}$1.1 & 177.397 & 22.396007 & 1.4888 & 22.46 $\pm$ 0.01 & 3.7$\pm$0.1\\
  $^{+}$1.2 & 177.39941 & 22.397438 & 1.4888 & 23.39 $\pm$ 0.01 & 4.1$\pm$0.1\\
  $^{+}$1.3 & 177.40341 & 22.402426 & 1.4888 & 22.73 $\pm$ 0.01 & 9.7$\pm$0.3\\
  2.1 & 177.40243 & 22.389739 & 1.894 & 26.46 $\pm$ 0.02 & 4.6$\pm$0.1\\
  2.2 & 177.40607 & 22.392484 & 1.894 & 24.4 $\pm$ 0.01 & $>$20\\
  2.3 & 177.40657 & 22.392881 & 1.894 & 24.49 $\pm$ 0.01 & 18.5$\pm$1.6\\
  3.1 & 177.39076 & 22.39984 & 3.128 & 23.36 $\pm$ 0.0 & 10.5$\pm$0.4\\
  3.2 & 177.39272 & 22.403074 & 3.128 & 22.77 $\pm$ 0.0 & 10.3$\pm$0.4\\
  3.3 & 177.40129 & 22.407182 & 3.128 & 24.01 $\pm$ 0.01 & 4.3$\pm$0.1\\
  4.1 & 177.39301 & 22.396826 & 2.95 & 25.41 $\pm$ 0.01 & -- \\
  4.2 & 177.3944 & 22.400729 & 2.95 & -- & 7.4$\pm$0.2\\
  4.3 & 177.40419 & 22.40612 & 2.95 & 25.96 $\pm$ 0.03 & 3.4$\pm$0.1\\
  5.1 & 177.39976 & 22.393062 & 2.79 & 25.15 $\pm$ 0.01 & 15.5$\pm$0.7\\
  5.2 & 177.40111 & 22.393824 & 2.79 & 25.01 $\pm$ 0.01 & 12.0$\pm$0.5\\
  5.3 & 177.40794 & 22.403538 & 2.79 & 26.12 $\pm$ 0.02 & 4.3$\pm$0.1\\
  6.1 & 177.39972 & 22.392545 & 2.66 $\pm$ 0.02 & 26.37 $\pm$ 0.03 & 9.0$\pm$0.3\\
  6.2 & 177.40181 & 22.393858 & -- & 26.4 $\pm$ 0.02 & 8.1$\pm$0.3\\
  6.3 & 177.40804 & 22.402505 & -- & 27.41 $\pm$ 0.06 & 4.7$\pm$0.1\\
  7.1 & 177.39895 & 22.391332 & 2.79 $\pm$ 0.02 & 25.87 $\pm$ 0.02 & 4.5$\pm$0.1\\
  7.2 & 177.40339 & 22.394269 & -- & 26.16 $\pm$ 0.02 & 4.6$\pm$0.1\\
  7.3 & 177.40759 & 22.401243 & -- & 26.3 $\pm$ 0.03 & 4.2$\pm$0.1\\
  8.1 & 177.39849 & 22.394351 & 2.81 $\pm 0.02$ & 26.12 $\pm$ 0.02 & $>$20\\
  8.2 & 177.39978 & 22.395055 & -- & 24.7 $\pm$ 0.04 & 15.1$\pm$0.6\\
  8.3 & 177.40709 & 22.40472 & -- & 26.03 $\pm$ 0.02 & 3.2$\pm$0.1\\
  9.1 & 177.40515 & 22.426221 & 0.981 & 24.81 $\pm$ 0.01 & 1.7$\pm$0.1\\
  9.2 & 177.40387 & 22.427217 & 0.981 & 24.57 $\pm$ 0.01 & 4.9$\pm$1.3\\
  9.3 & 177.40323 & 22.427221& 0.981 & 24.14 $\pm$ 0.0 & 2.9$\pm$0.3\\
  9.4 & 177.40365 & 22.426408 & 0.981 & 25.11 $\pm$ 0.01 & 3.4$\pm$0.3\\
  10.1 & 177.40447 & 22.425508 & 1.34 $\pm$ 0.01 & 25.99 $\pm$ 0.01 & 3.0$\pm$0.2\\
  10.2 & 177.40362 & 22.425629 & -- & 26.09 $\pm$ 0.01 & 2.2$\pm$0.1\\
  10.3 & 177.4022 & 22.426611 & -- & 26.5 $\pm$ 0.02 & 1.8$\pm$0.1\\
  13.1 & 177.4037 & 22.397787 & 1.28 $\pm$ 0.01 & 25.87 $\pm$ 0.03 & $>$20 \\
  13.2 & 177.40282 & 22.396656 & -- & 26.14 $\pm$ 0.02 & 11.9$\pm$0.6\\
  13.3 & 177.40003 & 22.393857 & -- & 25.78 $\pm$ 0.02 & 5.3$\pm$0.1\\
  14.1 & 177.39166 & 22.403504 & 3.50 $\pm$ 0.06 & 27.06 $\pm$ 0.03 & 13.4$\pm$0.7\\
  14.2 & 177.39084 & 22.402624 & -- & 27.13 $\pm$ 0.03 & $>$20\\
  15.1 & 177.40922 & 22.387695 & 3.58 $\pm$ 0.08 & 26.57 $\pm$ 0.03 & 7.5$\pm$0.9\\
  15.2 & 177.41034 & 22.388745 & -- & 25.86 $\pm$ 0.02 & $>$20\\
  15.3 & 177.40624 & 22.385349 & -- & 27.19 $\pm$ 0.04 & 3.8$\pm$0.1\\
  16.1 & 177.40971 & 22.387662 & 2.65 $\pm$ 1.45 & 27.19 $\pm$ 0.04 & $>$20\\
  16.2 & 177.40989 & 22.387828 & -- & 27.34 $\pm$ 0.04 & $>$20\\
  17.1 & 177.40994 & 22.387232 & 6.28 $\pm$ 0.17 & 28.02 $\pm$ 0.06 & 5.5$\pm$0.4\\
  17.2 & 177.41124 & 22.388457 & -- & 28.14 $\pm$ 0.07 & 15.2$\pm$1.1\\
  17.3 & 177.40658 & 22.384483 & -- & 28.46 $\pm$ 0.08 & 3.6$\pm$0.1\\
  18.1 & 177.40959 & 22.38666 & 7.76 $\pm$ 0.16 & 28.51 $\pm$ 0.23 & 3.4$\pm$0.2\\
  18.2 & 177.41208 & 22.389057 & -- & -- & 8.3$\pm$0.3\\
  18.3 & 177.40669 & 22.384319 & -- & -- & 3.6$\pm$0.1\\
  21.1 & 177.39284 & 22.41287 & 2.48 $\pm$ 0.04 & 26.38 $\pm$ 0.02 & $>$20\\
  21.2 & 177.39353 & 22.413083 & -- & 22.52 $\pm$ 0.06 & $>$20\\
  21.3 & 177.39504 & 22.412686 & -- & 27.5 $\pm$ 0.04 & 14.6$\pm$1.1\\
  22.1 & 177.40402 & 22.3929 & 3.216 & 27.86 $\pm$ 0.05 & 5.0$\pm$0.2\\
  22.2 & 177.40906 & 22.400233 & 3.216 & 27.85 $\pm$ 0.05 & 3.9$\pm$0.1\\
  22.3 & 177.40016 & 22.39015 & 3.216 & 27.57 $\pm$ 0.05 & 4.1$\pm$0.1\\
  26.1 & 177.40475 & 22.425978 & 1.49 $\pm$ 0.03 & 26.87 $\pm$ 0.03 & 3.5$\pm$0.4\\
  26.2 & 177.40361 & 22.426078 & -- & 26.44 $\pm$ 0.03 & 4.0$\pm$0.5\\
  26.3 & 177.40274 & 22.426936 & -- & 26.7 $\pm$ 0.02 & 2.5$\pm$0.1\\
  29.1 & 177.40799 & 22.389056 & 2.76 $\pm$ 0.05 & 27.99 $\pm$ 0.07 & 10.7$\pm$2.0\\
  29.2 & 177.40907 & 22.390406 & -- & 27.55 $\pm$ 0.04 & 9.2$\pm$0.4\\
  29.3 & 177.40451 & 22.386702 & -- & 28.56 $\pm$ 0.08 & 4.0$\pm$0.1\\
  31.1 & 177.40215 & 22.396747 & 2.78 $\pm$ 0.03 & 26.86 $\pm$ 0.03 & 2.3$\pm$0.1\\
  31.2 & 177.39529 & 22.391833 & -- & 26.2 $\pm$ 0.02 & 3.2$\pm$0.1\\
  31.3 & 177.40562 & 22.402439 & -- & 26.1 $\pm$ 0.02 & 4.2$\pm$0.1\\
  34.1 & 177.4082 & 22.388116 & 3.42 $\pm$ 0.08 & 27.28 $\pm$ 0.04 & 4.3$\pm$0.5\\
  34.2 & 177.41037 & 22.390621 & -- & 27.35 $\pm$ 0.05 & 6.7$\pm$0.2\\
  34.3 & 177.40518 & 22.386031 & -- & 27.66 $\pm$ 0.06 & 4.0$\pm$0.1\\
\hline 
\hline
$^{\ast}$1002.2 & 177.39701 & 22.396 & 1.4888 & -- & -- \\
$^{\ast}$1002.3 & 177.39943 & 22.397424 & 1.4888 & -- & -- \\
$^{\ast}$1002.1 & 177.40343 & 22.402419 & 1.4888 & -- & -- \\
$^{\ast}$1003.2 & 177.39815 & 22.396344 & 1.4888 & -- & -- \\
$^{\ast}$1003.3 & 177.39927 & 22.39683 & 1.4888 & -- & -- \\
$^{\ast}$1003.1 & 177.40384 & 22.402564 & 1.4888 & -- & -- \\
$^{\ast}$1004.2 & 177.39745 & 22.396397 & 1.4888 & -- & -- \\
$^{\ast}$1004.3 & 177.39916 & 22.397214 & 1.4888 & -- & -- \\
$^{\ast}$1004.1 & 177.40359 & 22.402653 & 1.4888 & -- & -- \\
$^{\ast}$1006.2 & 177.39712 & 22.396717 & 1.4888 & -- & -- \\
$^{\ast}$1006.3 & 177.39812 & 22.398247 & 1.4888 & -- & -- \\
$^{\ast}$1006.4 & 177.39878 & 22.397625 & 1.4888 & -- & -- \\
$^{\ast}$1006.1 & 177.40338 & 22.402867 & 1.4888 & -- & -- \\
$^{\ast}$1007.2 & 177.39697 & 22.396636 & 1.4888 & -- & -- \\
$^{\ast}$1007.3 & 177.39782 & 22.398464 & 1.4888 & -- & -- \\
$^{\ast}$1007.4 & 177.39882 & 22.397711 & 1.4888 & -- & -- \\
$^{\ast}$1007.1 & 177.40329 & 22.402831 & 1.4888 & -- & -- \\
$^{\ast}$1008.2 & 177.39698 & 22.396553 & 1.4888 & -- & -- \\
$^{\ast}$1008.3 & 177.39793 & 22.398418 & 1.4888 & -- & -- \\
$^{\ast}$1008.4 & 177.39889 & 22.397639 & 1.4888 & -- & -- \\
$^{\ast}$1008.1 & 177.40331 & 22.402786 & 1.4888 & -- & -- \\
$^{\ast}$1009.2 & 177.397 & 22.396444 & 1.4888 & -- & -- \\
$^{\ast}$1009.1 & 177.399 & 22.397625 & 1.4888 & -- & -- \\
$^{\ast}$1010.2 & 177.39694 & 22.396394 & 1.4888 & -- & -- \\
$^{\ast}$1010.1 & 177.39904 & 22.397611 & 1.4888 & -- & -- \\
$^{\ast}$1011.2 & 177.39701 & 22.396197 & 1.4888 & -- & -- \\
$^{\ast}$1011.3 & 177.39922 & 22.397472 & 1.4888 & -- & -- \\
$^{\ast}$1011.1 & 177.40337 & 22.402556 & 1.4888 & -- & -- \\
$^{\ast}$1015.2 & 177.39672 & 22.395372 & 1.4888 & -- & -- \\
$^{\ast}$1015.3 & 177.39975 & 22.397489 & 1.4888 & -- & -- \\
$^{\ast}$1015.4 & 177.40012 & 22.397203 & 1.4888 & -- & -- \\
$^{\ast}$1015.1 & 177.40325 & 22.402008 & 1.4888 & -- & -- \\
$^{\ast}$1016.2 & 177.39688 & 22.396211 & 1.4888 & -- & -- \\
$^{\ast}$1016.3 & 177.39918 & 22.397589 & 1.4888 & -- & -- \\
$^{\ast}$1016.1 & 177.40327 & 22.402581 & 1.4888 & -- & -- \\
$^{\ast}$1018.2 & 177.39723 & 22.396208 & 1.4888 & -- & -- \\
$^{\ast}$1018.3 & 177.39933 & 22.397303 & 1.4888 & -- & -- \\
$^{\ast}$1018.1 & 177.40354 & 22.402533 & 1.4888 & -- & -- \\
$^{\ast}$1019.2 & 177.39661 & 22.396308 & 1.4888 & -- & -- \\
$^{\ast}$1019.3 & 177.39777 & 22.398783 & 1.4888 & -- & -- \\
$^{\ast}$1019.4 & 177.39867 & 22.398219 & 1.4888 & -- & -- \\
$^{\ast}$1019.5 & 177.39899 & 22.397869 & 1.4888 & -- & -- \\
$^{\ast}$1019.1 & 177.40303 & 22.402681 & 1.4888 & -- & -- \\
$^{\ast}$1020.2 & 177.39708 & 22.395722 & 1.4888 & -- & -- \\
$^{\ast}$1020.1 & 177.40353 & 22.402236 & 1.4888 & -- & -- \\
$^{\ast}$1021.2 & 177.39689 & 22.395761 & 1.4888 & -- & -- \\
$^{\ast}$1021.3 & 177.39954 & 22.397483 & 1.4888 & -- & -- \\
$^{\ast}$1021.4 & 177.39996 & 22.397094 & 1.4888 & -- & -- \\
$^{\ast}$1021.1 & 177.40336 & 22.402289 & 1.4888 & -- & -- \\
$^{\ast}$1022.2 & 177.39717 & 22.396508 & 1.4888 & -- & -- \\
$^{\ast}$1022.3 & 177.39895 & 22.397503 & 1.4888 & -- & -- \\
$^{\ast}$1022.1 & 177.40342 & 22.402764 & 1.4888 & -- & -- \\
$^{\ast}$1024.2 & 177.39809 & 22.395853 & 1.4888 & -- & -- \\
$^{\ast}$1024.3 & 177.39993 & 22.396714 & 1.4888 & -- & -- \\
$^{\ast}$1024.1 & 177.40379 & 22.402193 & 1.4888 & -- & -- \\
$^{\ast}$1026.2 & 177.39798 & 22.396011 & 1.4888 & -- & -- \\
$^{\ast}$1026.3 & 177.39981 & 22.39676 & 1.4888 & -- & -- \\
$^{\ast}$1026.1 & 177.40379 & 22.402317 & 1.4888 & -- & -- \\
$^{\ast}$1050.2 & 177.39746 & 22.395653 & 1.4888 & -- & -- \\
$^{\ast}$1050.3 & 177.39761 & 22.395778 & 1.4888 & -- & -- \\
$^{\ast}$1050.4 & 177.39775 & 22.395217 & 1.4888 & -- & -- \\
$^{\ast}$1050.5 & 177.39818 & 22.395681 & 1.4888 & -- & -- \\
$^{\ast}$1050.6 & 177.40006 & 22.396691 & 1.4888 & -- & -- \\
$^{\ast}$1050.1 & 177.40376 & 22.402033 & 1.4888 & -- & -- \\
$^{\ast}$1052.2 & 177.3973 & 22.395383 & 1.4888 & -- & -- \\
$^{\ast}$1052.3 & 177.39792 & 22.395725 & 1.4888 & -- & -- \\
$^{\ast}$1052.4 & 177.39803 & 22.395239 & 1.4888 & -- & -- \\
$^{\ast}$1052.5 & 177.39817 & 22.395478 & 1.4888 & -- & -- \\
$^{\ast}$1052.6 & 177.40016 & 22.396758 & 1.4888 & -- & -- \\
$^{\ast}$1052.1 & 177.4037 & 22.401947 & 1.4888 & -- & -- \\
$^{\ast}$1192.2 & 177.39664 & 22.396236 & 1.4888 & -- & -- \\
$^{\ast}$1192.3 & 177.39796 & 22.398689 & 1.4888 & -- & -- \\
$^{\ast}$1192.4 & 177.39904 & 22.397833 & 1.4888 & -- & -- \\
$^{\ast}$1192.1 & 177.40305 & 22.402631 & 1.4888 & -- & -- \\
$^{\ast}$1211.2 & 177.39699 & 22.395628 & 1.4888 & -- & -- \\
$^{\ast}$1211.1 & 177.40346 & 22.402172 & 1.4888 & -- & -- \\
$^{\ast}$1222.2 & 177.39698 & 22.396081 & 1.4888 & -- & -- \\
$^{\ast}$1222.3 & 177.39938 & 22.397461 & 1.4888 & -- & -- \\
$^{\ast}$1222.1 & 177.4034 & 22.402461 & 1.4888 & -- & -- \\
\hline
\end{supertabular}
\end{center}



\label{lastpage}

\end{document}